\documentclass{gGAF2e}

\usepackage{amsbsy,amssymb,amsmath,bm}
\usepackage{graphicx}
\usepackage{subfigure}
\usepackage{multiequations}

\usepackage{natbib}

\begin{document}
\doi{10.1080/03091920xxxxxxxxx} \issn{1029-0419} \issnp{0309-1929} \jvol{00} \jnum{00} \jyear{2011} \jmonth{August}

\markboth{Wei, Jackson and Hollerbach}{Kinematic dynamo action in spherical Couette flow}

\title{Kinematic dynamo action in spherical Couette flow}

\author{Xing Wei${\dag}$$^{\ast}$,\thanks{$^\ast$Corresponding author. Email: xing.wei@erdw.ethz.ch \vspace{6pt}} Andrew Jackson${\dag}$ and Rainer Hollerbach${\dag}$${\ddag}$\\
\vspace{6pt} ${\dag}$Institut f\"ur Geophysik, ETH, Z\"urich 8092, Switzerland\\ ${\ddag}$Department of Applied Mathematics, University of Leeds, Leeds LS2 9JT, U.K\\
\vspace{6pt}\received{v1.0 released June 2011}}

\maketitle

\begin{abstract}
We investigate numerically kinematic dynamos driven by flow of electrically conducting fluid in the shell between two concentric differentially rotating spheres, a configuration normally referred to as spherical Couette flow. We compare between axisymmetric (2D) and fully three dimensional flows, between low and high global rotation rates, between prograde and retrograde differential rotations, between weak and strong nonlinear inertial forces, between insulating and conducting boundaries, and between two aspect ratios. The main results are as follows. Azimuthally drifting Rossby waves arising from the destabilisation of the Stewartson shear layer are crucial to dynamo action. Differential rotation and helical Rossby waves combine to contribute to the spherical Couette dynamo. At a slow global rotation rate, the direction of differential rotation plays an important role in the dynamo because of different patterns of Rossby waves in prograde and retrograde flows. At a rapid global rotation rate, stronger flow supercriticality (namely the difference between the differential rotation rate of the flow and its critical value for the onset of nonaxisymmetric instability) facilitates the onset of dynamo action. A conducting magnetic boundary condition and a larger aspect ratio both favour dynamo action.
\end{abstract}\bigskip

\bigskip

\section{Introduction}

Dynamo action is believed to generate magnetic fields in the universe \citep[e.g.][]{Ruediger_Hollerbach}. In dynamo action, the motion of conducting fluid produces a magnetic field through the effect of electromagnetic induction while the generated magnetic field gives a back-reaction on the fluid motion, and so the dynamo is a coupled nonlinear system. A simplified model is the kinematic dynamo in which the back-reaction of the field on the flow is neglected. Beginning with the seminal work of \citet{Bullard}, kinematic dynamo theory is a classic test-bed for discovering the efficiency of flows in generating magnetic field. The subject has already been extensively studied \citep{Roberts,Gubbins1,Gubbins2} and is most recently reviewed by \citet{Gubbins}. In the kinematic dynamo the fluid flow can either be prescribed \citep[e.g.][]{Galloway_Proctor} or be a solution to the Navier-Stokes (N-S) equation of fluid motion subject to a given forcing \citep[e.g.][]{Livermore}. In the latter case of prescribed forcing the nonlinear interaction between flow and field is removed by dropping the Lorentz force in the N-S equation. Because in the kinematic dynamo the magnetic field has no back-reaction on the fluid flow, the magnetic induction equation is linear and thus the magnetic field does not saturate but grows.

In the geodynamo driven by convective fluid motion in the Earth's core, it is plausible that a large scale differential rotation in the zonal flow emerges because of angular momentum transfer through convection \citep{Busse,Manneville,Aurnou}, i.e. the so-called thermal wind, or through turbulent Reynolds stress \citep{Ruediger}, such that the solid inner core rotates at a different rate relative to the mantle due to the viscous coupling exerted on the inner core. This large scale differential rotation has been supported by the recent geodynamo simulations at a very low Ekman number \citep{Sakuraba} and the super-rotation of the solid inner core by some seismic data \citep{Zhang_Song} (though it still remains controversial \citep{Souriau}). This differential rotation is important in dynamo action, e.g. through the $\omega$ effect that shears poloidal magnetic field lines to create toroidal magnetic field lines. Many simulations of the geodynamo driven by thermal and compositional convection have been carried out in the geometry of a spherical shell \citep{Roberts_Glatzmaier, Wicht_Tilgner, Jones}. However, the convectively driven dynamo is very complicated, and therefore a simplified dynamo model is proposed, i.e. the spherical Couette dynamo, in which the dynamo is driven by an imposed mechanical force arising from the differential rotation of the inner and outer spheres. This simple dynamo model captures one of the important ingredients in dynamo action, namely differential rotation. The need to understand dynamo action at realistic values of the magnetic Prandtl number has spawned a number of liquid metal dynamo experiments using differential rotation in a spherical geometry, examples of which are in Grenoble \citep{Grenoble} and Maryland \citep{Lathrop}. These experiments motivate the present study.

Flow in a spherical shell driven by differentially rotating inner and outer spheres is called spherical Couette flow. Spherical Couette flow at an infinitesimally differential rotation rate was analytically studied by \citet{Stewartson}. He showed that the fluid outside the tangent cylinder, a cylinder parallel to the rotational axis and touching the inner sphere, rotates at the rate of the outer sphere due to it experiencing a homogeneous boundary condition, whereas the fluid inside the tangent cylinder rotates at an intermediate rate between the rotation rates of inner and outer spheres due to it experiencing an inhomogeneous boundary condition. Consequently a shear layer in the angular velocity emerges at the tangent cylinder, which is called the Stewartson layer. Subsequently, the situation of finite differential rotation rates was numerically studied by Hollerbach and coworkers. \citet{JFM} calculated the axisymmetric basic state and the nonaxisymmetric linear stability of spherical Couette flow, and pointed out that the instability of Stewartson flow is asymmetric in the direction of differential rotation, namely retrograde flow is more stable than prograde flow. \citet{TCFD} \citep[see also][]{Schaeffer_Cardin1} calculated the nonlinear saturation of instability, and showed that the destabilisation of the Stewartson layer triggers azimuthally drifting Rossby waves due to the curvature of the spherical geometry, the so-called $\beta$ effect \citep{Pedlosky}.

In addition to nonmagnetic spherical Couette flow, the electrically conducting spherical Couette flow with an imposed field is called the magnetic spherical Couette flow. The axisymmetric basic state of magnetic spherical Couette flow was studied analytically by \citet{Starchenko}, numerically by
%  Dormy and coworkers 
\citet{Dormy1,Dormy2}
% \citeyearpar{Dormy1,Dormy2} 
and 
%  Hollerbach
\citet{LNP,PRSA1,EJMB},
%  and coworkers \citeyearpar{LNP,PRSA1,EJMB}, 
and experimentally in Maryland \citep{Lathrop1} and Grenoble \citep{Grenoble1,Grenoble2}. When the imposed field is uniformly aligned with the rotation axis it enhances the shear in the Stewartson layer, however, when the imposed field is not parallel to the rotation axis (say, dipolar or quadrupolar) the increasing imposed field alters the Stewartson layer until a new curved shear layer emerges, called the Shercliff layer. The nonaxisymmetric instability in magnetic spherical Couette flow was then studied with an imposed field parallel to the rotation axis \citep{PRE,PRSA2}. It is found that the imposed field can suppress the asymmetry of instabilities between prograde and retrograde flows in the nonmagnetic Stewartson layer.

We now come back to the Couette dynamo problem. Dynamo action induced by Taylor-Couette flow in a cylindrical geometry has been studied by \citet{Willis_Barenghi} in both linear and nonlinear regimes, in which shear and Taylor vortices are combined to contribute to dynamo action. However, the two geometries are quite different; the cylindrical problem has neither Stewartson layers nor a $\beta$ effect to trigger Rossby waves, whereas the spherical problem has no Taylor vortices.

\citet{Schaeffer_Cardin} studied the spherical kinematic dynamo in a particular form of rapidly rotating flow. There is no inner sphere in their spherical geometry but the outer sphere is split into two parts along the tangent cylinder, the region outside the tangent cylinder rotating at the global rate and the top and bottom boundaries inside rotating at a different rate. In this configuration the Stewartson shear layer also emerges because of the angular velocity jump and the destabilisation of the shear layer triggers Rossby waves. A quasi-geostrophic model for the flow was used, in which the cylindrical radial and azimuthal velocities are invariant along the rotation axis ($z$ direction) but the axial velocity is linearly dependent on the $z$ coordinate, such that a regime of extremely low viscosity is numerically achievable (very low Ekman and moderately small magnetic Prandtl numbers). In purely hydrodynamic calculations they found that for prograde flow Rossby waves tend to spread out of the shear layer whereas those for retrograde flow tend to be confined in the vicinity of the shear layer. In the kinematic dynamo calculations they found that retrograde flow slightly reduces the dynamo threshold. The spatial structure of Rossby waves is helical and it is well known that a helical wave can induce the alpha effect \citep{Parker,Moffatt}, the effect that twists toroidal magnetic field lines to create poloidal magnetic field lines. Therefore this dynamo was interpreted with the $\alpha$-$\omega$ mechanism in which the $\omega$ effect is induced by the shear and the $\alpha$ effect by Rossby waves. Moreover, in their calculations both dipolar and quadrupolar fields were generated.

\citet{Guervilly_Cardin} then studied the nonlinear spherical Couette dynamo in which the nonlinear interaction of fluid flow and magnetic field is accounted for by retaining the Lorentz force in the N-S equation. They calculated both the case of a stationary outer sphere without global rotation and the case of a rotating outer sphere with global rotation. They found that the resulting axisymmetric flow cannot generate a dynamo until nonaxisymmetric hydrodynamical instabilities are excited, and that the presence of global rotation facilitates the onset of dynamo action. Without global rotation the critical magnetic Prandtl number $Pm$ is of the order of unity, which may be translated into a magnetic Reynolds number $Rm$ to be of the order of a few thousand. However, with global rotation the critical $Pm$ is reduced by half, and retrograde flow (negative Rossby number) is better for dynamo action. More interestingly, a dynamo window in which $Rm\sim300$ appears in the regime of high Ekman number $E=10^{-3}$ and negative Rossby number $-2\le Ro\le -1.5$. This dynamo window was physically interpreted as the enhancement of dynamo action by the shear layer.

Since the nonlinear spherical Couette dynamo is complicated to understand, the kinematic spherical Couette dynamo is numerically investigated in this paper, in which the interaction of fluid flow and magnetic field is decoupled by dropping the Lorentz force. We do not study the case of a stationary outer sphere without global rotation but instead focus on the case of a rotating outer sphere with global rotation which is relevant to the Earth. Although this kinematic dynamo is simple, it consists of the two most important ingredients in dynamo action, namely the $\omega$ effect induced by differential rotation and the $\alpha$ effect induced by helical Rossby waves. In what follows we describe our numerical calculations of kinematic dynamos in spherical Couette flow. In section 2 we show the governing equations and introduce the numerical method, in section 3 we discuss the results in detail, and in section 4 we draw some conclusions.

\section{Governing equations and numerical method}

Following \citet{JFM}, the dimensionless Navier-Stokes equation of fluid motion is
\begin{equation}\label{eq:ns}
\frac{\partial\bm U}{\partial t} + |Ro|\bm U\bm\cdot\bm\nabla\bm U = -{\bm\nabla} p + E\nabla^2\bm U + 2\bm U\times\hat{\bm e}_z, \end{equation}
and the dimensionless magnetic induction equation is
\begin{equation}\label{eq:induction}
|Ro|^{-1}\frac{\partial\bm B}{\partial t} = \bm\nabla\times\left(\bm U\times\bm B\right) + Rm^{-1}\nabla^2\bm B.
\end{equation}
In (1) and (2), length is normalised with the spherical gap $L=r_o-r_i$, time with the inverse of global rotation rate $\varOmega^{-1}$, velocity with $|\varDelta\varOmega| L$ where $\varDelta\varOmega$ is the differential rotation rate, pressure with $\rho\varOmega|\varDelta\varOmega| L^2$ where $\rho$ is the fluid density, and magnetic field with $\sqrt{\rho\mu\varOmega|\varDelta\varOmega|}L$ where $\mu$ is the vacuum magnetic permeability. It should be noted that the absolute value of differential rotation $|\varDelta\varOmega|$ but not $\varDelta\varOmega$ itself is used in the normalisation. The reason for this is to ensure that the dimensional and dimensionless velocities are always parallel rather than anti-parallel.

In the governing equations there are three dimensionless parameters, $E$, $Ro$ and $Rm$. The Ekman number
\begin{multiequations}\label{eq:parameters}
\begin{equation}\label{eq:ek}
\singleequation
E=\frac{\nu}{\varOmega(r_o-r_i)^2}
\end{equation}
measures the global rotation, where $\nu$ is the fluid viscosity. The Rossby number
\begin{equation}\label{eq:ro}
\singleequation
Ro=\frac{\varDelta\varOmega}{\varOmega}
\end{equation}
measures the differential rotation. The magnetic Reynolds number
\begin{equation}\label{eq:rm}
\singleequation
Rm=\frac{|\varDelta\varOmega|(r_o-r_i)^2}{\eta}
\end{equation}
measures the electromagnetic induction effect, where $\eta$ is the magnetic diffusivity. Moreover, the Reynolds number $Re=|\varDelta\varOmega|L^2/\nu$ and the magnetic Prandtl number $Pm=\nu/\eta$ can be further deduced using these three numbers,
\begin{equation}
%   \label{eq:re}
Re = \frac{|Ro|}{E},
%   \end{equation}
%   \begin{equation}\label{eq:pm}
%   \singleequation
\qquad\qquad
Pm = \frac{Rm}{Re} = \frac{Rm\,E}{|Ro|}.
\end{equation}
\end{multiequations}

In addition to glocal Reynolds number $Re$ we can further define Reynolds number of $\omega$ effect $Re_\omega$ and Reynolds number of $\alpha$ effect $Re_\alpha$. $Re_\omega$ should be defined with the axisymmetric azimuthal
flow, and because in the spherical Couette flow the axisymmetric azimuthal energy is nearly the total
axisymmetric energy we define
\begin{multiequations}\label{eq:re_omega_alpha}
\begin{equation}\label{eq:re_omega}
\singleequation
Re_\omega = \sqrt{\frac{EK_0}{EK}}Re,
\end{equation}
where $EK$ is the total kinetic energy, $EK_0$ is the axisymmetric kinetic energy. $Re_\alpha$ should be defined with the nonaxisymmtric kinetic energy
\begin{equation}\label{eq:re_alpha}
\singleequation
Re_\alpha = \sqrt{\frac{EK_1}{EK}}Re,
\end{equation}
\end{multiequations}
where $EK_1$ is the nonaxisymmetric kinetic energy.

To achieve a Couette dynamo in laboratory experiments, $Rm$ should be above its critical value ($\sim10^2$) and $Pm$ in many conducting liquids is very small ($\sim10^{-6}$), such that either $|Ro|$ should be large enough or $E$ should be small enough (equation (\ref{eq:parameters}e)).
%   {eq:pm})). 
Therefore both the global and differential rotation rates as well as the length scale of fluid motion should be large enough, which requires strong mechanical driving.

For comparison between our kinematic spherical Couette dynamo and the nonlinear spherical Couette dynamo \citep{Guervilly_Cardin}, we list below the translations of parameters between our and their definitions,
\begin{multiequations}\label{eq:gc}
\begin{equation}
%  \label{eq:ek_gc}
\trippleequation
E=\frac{E_{gc}}{(1-\chi)^2},
\qquad\qquad
%   \end{equation}
%   \begin{equation}
%  \label{eq:ro_gc}
%   \singleequation
Ro=\frac{Ro_{gc}}{\chi},
\qquad\qquad
%   \end{equation}
%   and
%   \begin{equation}
%  \label{eq:rm_gc}
%   \singleequation
Rm=\frac{(1-\chi)^2}{\chi}Rm_{gc},
\end{equation}
\end{multiequations}
where the subscript ``gc'' denotes the definition in \citet{Guervilly_Cardin} and $\chi$ is the aspect ratio $0.35$ in \citet{Guervilly_Cardin}.

The velocity boundary conditions are standard no-slip conditions in Couette flow,
\begin{multiequations}
\label{eq:u_bc}
\begin{equation}
\bm U=\mathbf 0 \hspace{3mm}\text{at}\hspace{3mm} r=r_o, \hspace{15mm} \bm U=r\sin\theta\frac{Ro}{|Ro|}\hat{\bm\phi} \hspace{3mm}\text{at}\hspace{3mm} r=r_i,
\end{equation}
\end{multiequations}
where $\hat{\bm\phi}$ is the azimuthal unit vector. In most of our numerical calculations $r_o$ and $r_i$ are set to be $1.5$ and $0.5$ in order to simulate the aspect ratio $1/3$ of the Earth's core, except in one calculation in which $r_o=2$ and $r_i=1$ are chosen to have a different aspect ratio $1/2$ for comparison with $1/3$. The magnetic boundary conditions in most of our numerical calculations are those appropriate to matching the field to an insulating exterior (both outside the outer sphere and inside the inner sphere), except in one calculation in which perfectly electrically conducting boundaries are employed for comparison with the insulating boundaries. The magnetic boundary conditions are then
\begin{multiequations}
\label{eq:b_bc}
\begin{equation}
\bm\nabla\times\bm B=\bm 0 \hspace{3mm}\text{for}\hspace{3mm}\text{i}, \hspace{15mm}
\hat{\bm r}\cdot\bm B=0 \hspace{3mm}\text{and}\hspace{3mm} \hat{\bm r}\times(\bm\nabla\times\bm B)=\bm 0 \hspace{3mm}\text{for}\hspace{3mm}\text{c},
\end{equation}
\end{multiequations}
where $\hat{\bm r}$ is the radial unit vector, and ``i'' denotes insulating boundary and ``c'' perfectly conducting boundary.

It should be also noted that although (1) and (2) are decoupled, the solution to (1) consists of time-dependent azimuthally drifting waves when $|Ro|$ is above its critical value, and therefore in the calculation of 3D flow-driven dynamos we need to solve (1) and (2) simultaneously.

The calculations are carried out with the pseudo-spectral code of \citet{IJNMF}. The toroidal-poloidal decomposition method is employed such that the divergence free condition of the incompressible fluid flow and of the magnetic field is automatically satisfied,
\begin{multiequations}
\label{eq:t_p}
\begin{equation}
\bm U=\bm\nabla\times(e\,\hat{\bm r})+\bm\nabla\times\bm\nabla\times(f\,\hat{\bm r}),\hspace{10mm}\bm B=\bm\nabla\times(g\,\hat{\bm r})+\bm\nabla\times\bm\nabla\times(h\,\hat{\bm r}).
\end{equation}
\end{multiequations}
The geometry is a spherical shell such that spherical harmonics are used. Take the toroidal scalar of velocity $e$ for example,
\begin{equation}\label{eq:spherical}
e\left(r,\theta,\phi,t\right)=\sum_{l,m}e_{lm}\left(r,t\right)\mathrm P_l^{|m|}(\cos\theta)e^{{\mathrm i}m\phi},
\end{equation}
where $\mathrm P_l^{|m|}$ is the associated Legendre polynomials of the degree $l$ and the order $m$. The Chebyshev polynomials are used in the radial direction,
\begin{multiequations}
\label{eq:radial}
\begin{equation}
e_{lm}\left(r,t\right)=\sum_{k}e_{klm}\left(t\right)\mathrm T_k(x),\hspace{13mm}x=(2r-r_o-r_i)/(r_o-r_i),
\end{equation}
\end{multiequations}
where $\mathrm T_k(x)$ is the Chebyshev polynomials. A 2nd order Runge-Kutta method is employed for time integration. In our calculations, Chebyshev polynomials with truncations as high as $170$ are used, and a degree and an order of spherical harmonics as high as, respectively, $170$ and $20$, are used. To search for the critical $Rm$ for dynamo action, we vary $Rm$ in steps of $50$, such that the error in the determination of the critical $Rm$ is within $50$, which is sufficiently small compared to the typical value of critical $Rm=\mathrm O(10^3)$ itself. For example, in the case $E=10^{-3.5}$ and $Ro=-1$, at $Rm=2600$ the magnetic energy decays whereas at $Rm=2650$ it grows, so we know that the critical $Rm$ is between $2600$ and $2650$.

In the magnetic induction equation (\ref{eq:induction}), the advection time scale for spherical Couette flow is of the order of unity and short, and therefore the time for the growth/decay rate of magnetic field to settle in is determined by magnetic diffusion time scale or $Rm/Ro$. $Rm$ and $Ro$ of the dynamo solutions are of the order of, respectively, $10^3$ and unity, such that it takes around one month to obtain one solution with the serial code which was used in our calculations.

\section{Results and discussions}

\begin{figure}[h]
\centering
\includegraphics[scale=0.5]{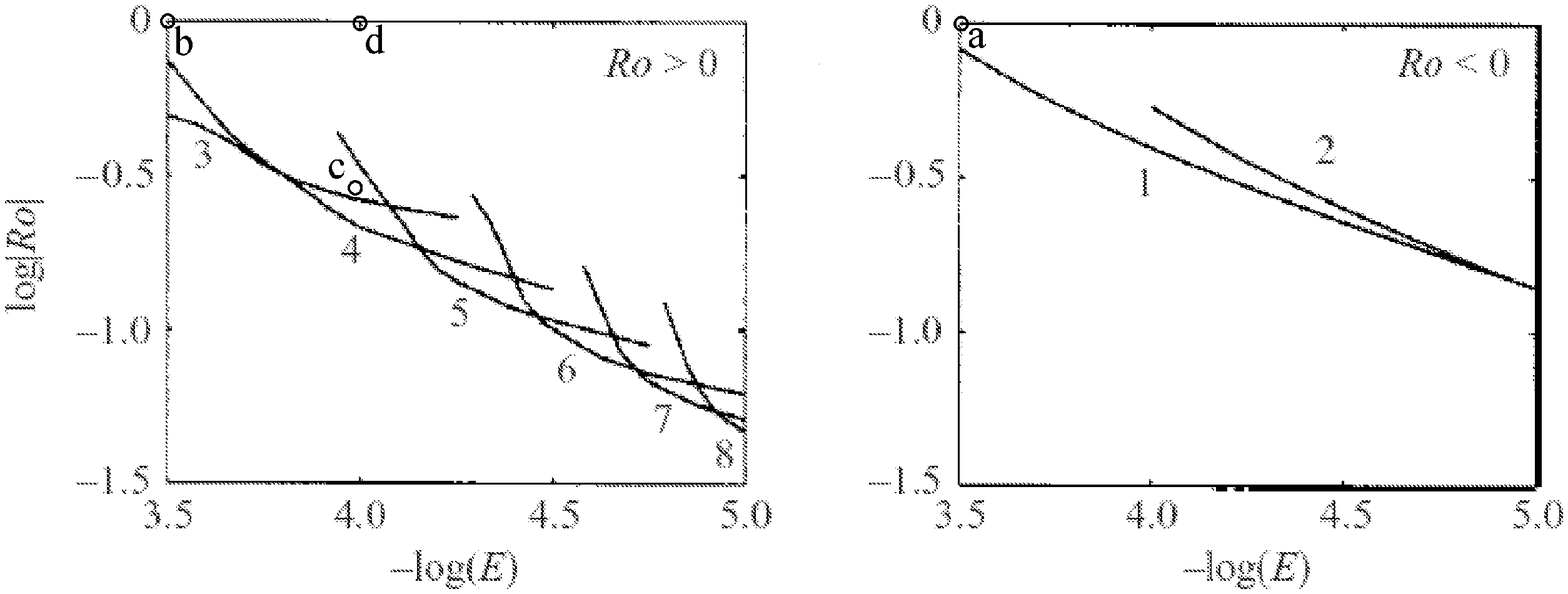}
\caption{The neutral stability curve of spherical Couette flow (figure 4 in \citep{JFM}, courtesy of Journal of Fluid Mechanics (Cambridge University Press)). $E=\nu/(\varOmega(r_o-r_i)^2)$, $Ro=\varDelta\varOmega/\varOmega$. Left panel is for prograde flow ($Ro>0$) and right panel retrograde flow ($Ro<0$). Integers beside the curves indicate the most unstable azimuthal modes. Letters a, b, c and d represent the first five solutions in Table \ref{tab:tab1}.}
\label{fig:fig0}
\end{figure}

In figure \ref{fig:fig0} we reproduce the results for linear stability calculations given in figure 4 of \citet{JFM}, showing the neutral stability curves for spherical Couette flow. The Ekman and Rossby numbers defined in figure \ref{fig:fig0} are identical to our definitions in equations (\ref{eq:parameters}a) and (\ref{eq:parameters}b). In our calculations, two values of Ekman number, $10^{-3.5}$ and $10^{-4}$, are chosen for comparison between low and high global rotation rates. At $E=10^{-3.5}$ the critical $Ro$ for the onset of Rossby waves is $Ro_c=+0.5$ for prograde flow and $Ro_c=-0.8$ for retrograde flow, while at $E=10^{-4}$ the critical $Ro$ is $Ro_c=+0.2$ for prograde flow and $Ro_c=-0.4$ for retrograde flow. We firstly try to use axisymmetric flows at $E=10^{-3.5}$ and $E=10^{-4}$ to produce a dynamo, namely $Ro$ numbers used in these calculations are below the neutral stability curves in figure \ref{fig:fig0} ($|Ro|<|Ro_c|$) such that no nonaxisymmetric instability occurs and flow is 2D. But unfortunately, regardless of whether we use a prograde or retrograde flow, all these axisymmetric flows fail to produce a dynamo (in the calculations we increase $Rm$ to $10^4$ but no dynamo occurs). So axisymmetric flow at this Ekman number regime is not favourable for dynamo action. This is the same as in the nonlinear Couette dynamo \citep{Guervilly_Cardin}.

We then increase $Ro$ slightly above its critical value to invoke the nonaxisymmetric components of flow which are azimuthally drifting Rossby waves arising from the destabilisation of the Stewartson layer \citep{TCFD}, and we achieve some dynamos. In what follows we show six illustrative dynamo solutions. Table 1 gives the parameters of the six dynamo solutions shown in the figures. The critical $Rm$ of these dynamo solutions is $Rm_c=\mathrm O(10^3)$ and the magnitude of the Rossby number is $|Ro|=\mathrm O(1)$ (see Table \ref{tab:tab1}). We calculate until $|Ro| \le 1$; this is a little more modest than the nonlinear dynamo calculations \citep{Guervilly_Cardin} in which $Ro$ translated to our definition reaches $+8.6$ for prograde flow and $-11.4$ for retrograde flow. According to equation (\ref{eq:parameters}e),
%  {eq:pm}), 
$Pm$ in our calculations is between $\mathrm O(0.1)$ and $\mathrm O(1)$, which is usually used in most simulations of convectively driven dynamo. In the first four solutions we employ insulating boundary conditions and an aspect ratio $r_i/r_o=1/3$ (letters a, b, c and d correspond to the circles in figure \ref{fig:fig0}). In the fifth solution we use perfectly electrically conducting boundary conditions and in the sixth $r_i/r_o=1/2$.

\begin{table}[h]
\centering
\setlength{\tabcolsep}{3pt}
\begin{tabular}{llllllllllllll} \hline
Figure & $E$ & $Ro$ & $Rm_c$ & $r_i/r_o$ & b.c. & $EK_t$ & $EM_t$ & $Re_\omega$ & $Re_\alpha$ & parity & $K$ & $L$ & $M$ \\ \hline
\ref{fig:fig1}, \ref{fig:fig1c} (a) & $10^{-3.5}$ & $-1$ & $2650$ & $1/3$ & i & 98.6 & 97.8 & 3132 & 439 & D & 50/50 & 50/50 & 5/5 \\
\ref{fig:fig2}, \ref{fig:fig2c} (b) & $10^{-3.5}$ & $+1$ & $5000$ & $1/3$ & i & 92.8 & 78.6 & 2897 & 1266 & Q & 70/80 & 70/80 & 10/20 \\
\ref{fig:fig3}, \ref{fig:fig3c} (c) & $10^{-4}$ & $+0.3$ & $6500$ & $1/3$ & i & 97.7 & 92.3 & 2898 & 775 & Q & 80/170 & 80/170 & 10/16 \\
\ref{fig:fig4}, \ref{fig:fig4c} (d) & $10^{-4}$ & $+1$ & $2000$ & $1/3$ & i & 85.9 & 79.7 & 8430 & 5380 & Q & 120/170 & 120/170 & 20/16 \\
\ref{fig:fig5a} \hspace{1mm} (a) & $10^{-3.5}$ & $-1$ & $2450$ & $1/3$ & c & 98.6 & 99.5 & 3132 & 439 & D & 50/50 & 50/50 & 5/5 \\
\ref{fig:fig6} & $10^{-3.5}$ & $-1$ & $1250$ & $1/2$ & i & 98.7 & 97.0 & 3099 & 625 & D & 50/50 & 50/50 & 5/5 \\ \hline
\end{tabular}
\vspace{5mm}
\caption{\large The parameters of six dynamo solutions. Column 1 gives the number of the figure illustrating the solutions, and letters a, b, c, and d correspond to the circles in figure \ref{fig:fig0}. $Rm_c$ in column 4 is an upper bound of the critical $Rm$. Column 6 shows the magnetic boundary condition, ``i'' denoting insulating boundary and ``c'' perfectly conducting boundary. Columns 7 and 8 show the percentage (\%) of, respectively, toroidal kinetic energy and toroidal magnetic energy. Columns 9 and 10 show the Reynolds numbers of, respectively, $\omega$ effect and $\alpha$ effect. Column 11 shows the parity of magnetic field, `D' denoting dipolar equatorial symmetry and `Q' quadrupolar equatorial symmetry. Columns 12, 13 and 14 show the spectral truncations of, respectively, Chebyshev polynomials, degree and order of spherical harmonics, number on the left of slash for flow $\bm U$ and number on the right for field $\bm B$.}
\label{tab:tab1}
\end{table}

The fact that a 2D flow at the Ekman number regime in our calculations cannot produce a dynamo but a 3D flow can indicates that azimuthally drifting Rossby waves are crucial in this configuration for dynamo action because Rossby waves with a helical spatial structure can induce an $\alpha$ effect of dynamo action \citep{Radler}, namely the motion of helical waves twists toroidal magnetic field lines to create poloidal magnetic field lines. Here we need to emphasise that our usage of the term $\alpha$ effect is a qualitative description of the helical nature of the flow, but is neither connected to a kinematically described $\alpha$ effect nor to an electromotive force that results from an average over small scale correlations in the velocity and magnetic fields, namely the mean field theory \citep{meanfield}. On the other hand, in all the dynamo solutions the toroidal kinetic energy of the zonal flow which is in a differential rotation occupies around $90\%$ of the total kinetic energy, and the toroidal magnetic energy occupies around $80\%$ or even higher of the total magnetic energy (see Table \ref{tab:tab1}). Therefore it is reasonable to conclude that these spherical Couette dynamos are $\alpha$-$\omega$ dynamos in which the $\omega$ effect is induced by differential rotation and the $\alpha$ effect is induced by azimuthally drifting Rossby waves.

\begin{figure}[h]
\centering
\subfigure[Contours of axisymmetric flow and field in the meridional plane. From left to right, angular velocity $U_\phi/(r\sin\theta)$, meridional circulation $\varPsi$ satisfying ${\bm\nabla}\times (\varPsi{\hat{\bm\phi}}/(r\sin\theta))=(U_r,U_\theta)$, toroidal field $B_\phi$ and lines of poloidal field $\chi$ satisfying ${\bm\nabla}\times(\chi{\hat{\bm\phi}}/(r\sin\theta))=(B_r,B_\theta)$. In the panels of angular velocity and toroidal field, solid lines represent west-east direction and dashed lines east-west. In the panels of meridional circulation and poloidal field, solid lines represent anti-clockwise direction and dashed lines clockwise. Contour lines are uniformly spaced with $10$ levels.]
{\includegraphics[scale=0.4]{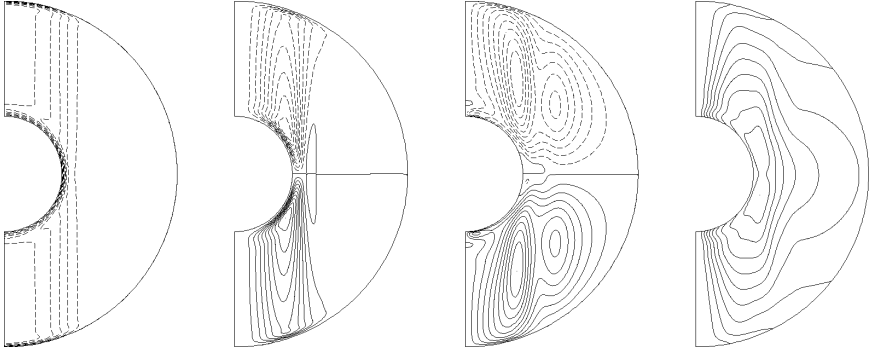}
\label{fig:fig1a}}
\subfigure[Contours of flow and field in the equatorial plane. From left to right, radial ($U_r$ and $B_r$), colatitude ($U_\theta$ and $B_\theta$) and azimuthal ($U_\phi$ and $B_\phi$) components of flow (top row) and field (bottom row). Solid lines represent positive value and dashed lines negative. Contour lines are uniformly spaced with $10$ levels.]
{\includegraphics[scale=0.4]{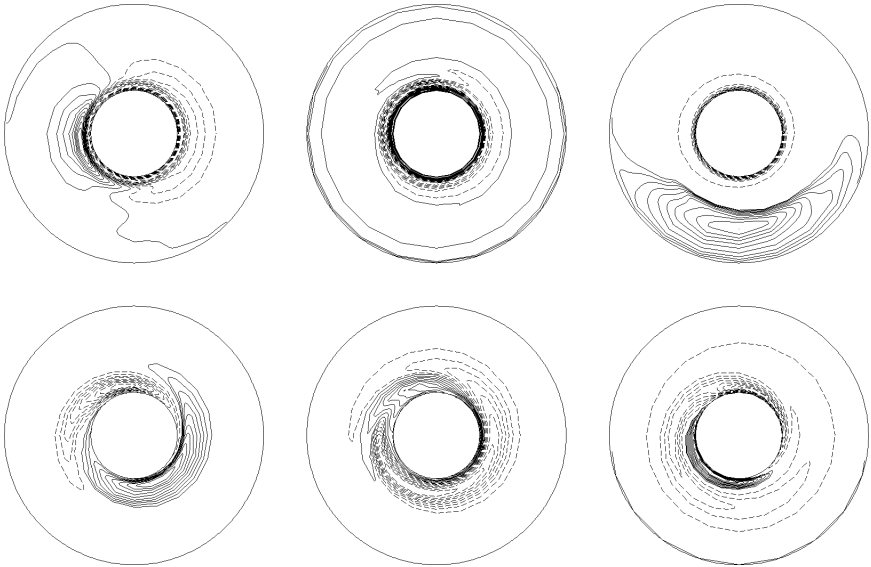}
\label{fig:fig1b}}
\caption{$E=10^{-3.5}$, $Ro=-1$ and $Rm_c=2650$.}
\label{fig:fig1}
\end{figure}

\begin{figure}[h]
\centering
\includegraphics[scale=0.9]{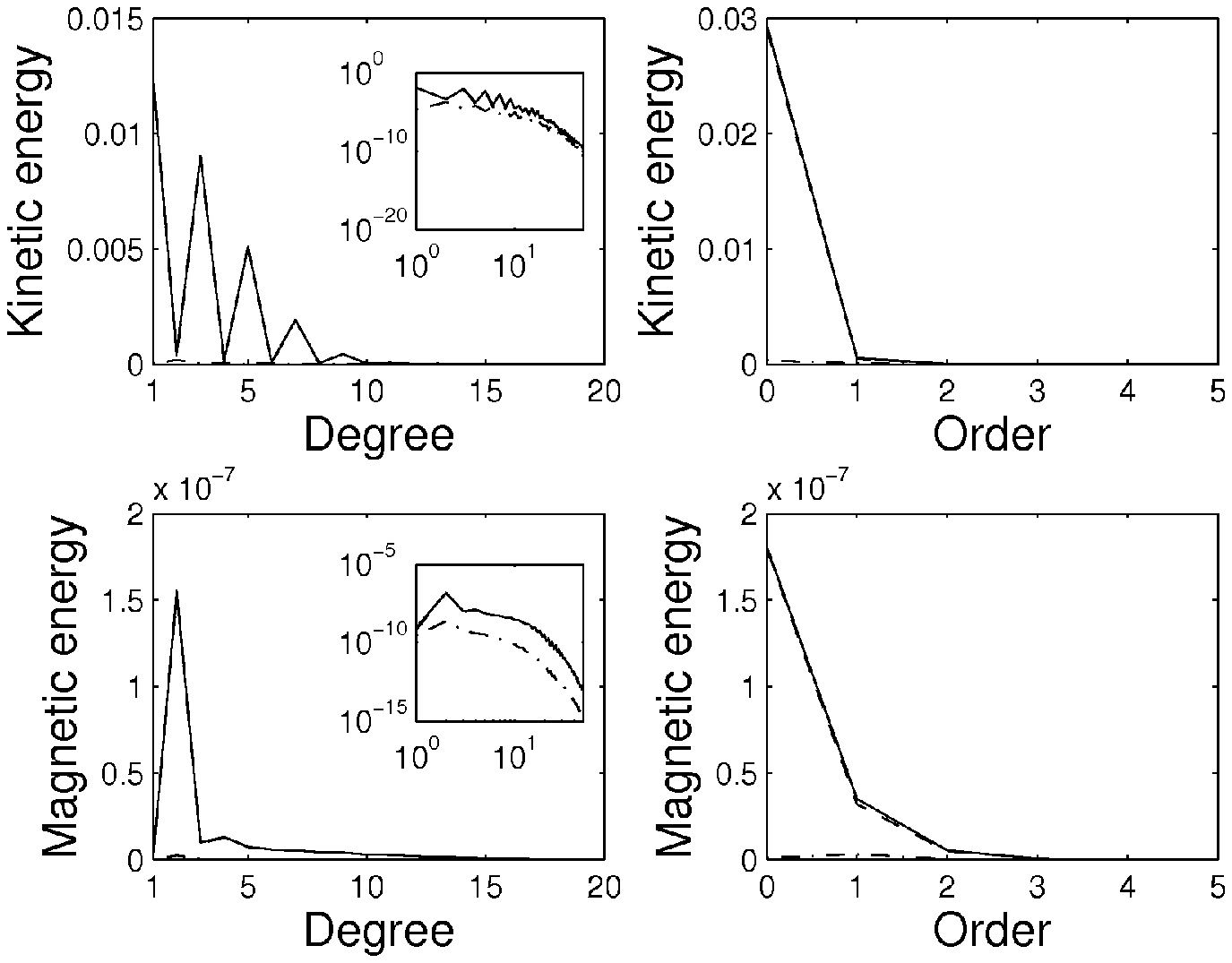}
\caption{Energy spectrum of flow and field for the solution in figure \ref{fig:fig1}. The top row shows the kinetic energy and the bottom row the magnetic energy. The left column shows the energy as a function of the degree $l$ of spherical harmonics and the right column the energy as a function of the order $m$. The large figures are plotted using a linear scale and the inset figures in the left column embedded in the large figures are plotted using a logarithmic scale. The solid lines represent the total energy, the dashed lines the toroidal energy and the dot-dashed lines the poloidal energy.}
\label{fig:fig1c}
\end{figure}

Figures \ref{fig:fig1} and \ref{fig:fig1c} show the dynamo solution at $E=10^{-3.5}$, $Ro=-1$ and $Rm_c=2650$. Figure \ref{fig:fig1a} shows the axisymmetric flow and field in the meridional plane and figure \ref{fig:fig1b} the flow and field distributions in the equatorial plane. The angular velocity and meridional circulation in figure \ref{fig:fig1a} exhibit the Stewartson layer and the process of Ekman pumping. The flow in the equatorial plane indicates that the Stewartson layer has already been destabilised at $Ro=-1$ and the $m=1$ mode of an azimuthally drifting wave has emerged, which is consistent with the linear stability calculations for spherical Couette flow (right panel in figure \ref{fig:fig0}). The poloidal magnetic field exhibits a dipolar symmetry about the equator and in the equatorial plane the field exhibits an $m=1$ mode.

To better understand this dynamo we diagnose some percentages of kinetic and magnetic energies. Figure \ref{fig:fig1c} shows the spectrum of kinetic and magnetic energies. The percentage of toroidal kinetic energy is $98.6\%$ (Table \ref{tab:tab1}) and the percentages of kinetic energies in $m=0$ and $m=1$ modes are, respectively, $98.1\%$ and $1.8\%$ (upper right panel in figure \ref{fig:fig1c}). Clearly the axisymmetric azimuthal flow is dominant and the differential rotation of this flow shears the poloidal field to create a strong toroidal field, i.e. the $\omega$ effect. This assertion is supported by the dominance of toroidal magnetic energy, namely that the percentage of toroidal magnetic energy is $97.8\%$ (Table \ref{tab:tab1}). The percentages of magnetic energies in the $m=0$ and $m=1$ modes are, respectively, $81.4\%$ and $15.6\%$ (lower right panel in figure \ref{fig:fig1c}). So this dynamo is an $\alpha$-$\omega$ dynamo, and the $\alpha$ effect is induced by Rossby waves which mainly consist of the $m=1$ mode (here we address again that the $\alpha$ effect does not connect to the mean field theory). A weak flow in the $m=1$ mode ($1.8\%$) can generate a relatively stronger poloidal field ($2.2\%$), so this Rossby-wave-induced $\alpha$ effect is efficient. The efficiency may be interpreted through the location of the $\alpha$ and $\omega$ effects. Figure \ref{fig:fig1b} shows that Rossby waves are concentrated in the vicinity of the inner sphere where the Stewartson layer is located and where there is a strong differential rotation to induce the $\omega$ effect. Therefore the $\alpha$ effect arising from Rossby waves and the $\omega$ effect arising from differential rotation work at the same location, which strengthens dynamo action. The efficiency resulting from location is discussed in detail by Gubbins and Gibbons \citeyearpar{Gubbins_Gibbons}.

\begin{figure}[h]
\centering
\subfigure[As in figure \ref{fig:fig1a}]
{\includegraphics[scale=0.4]{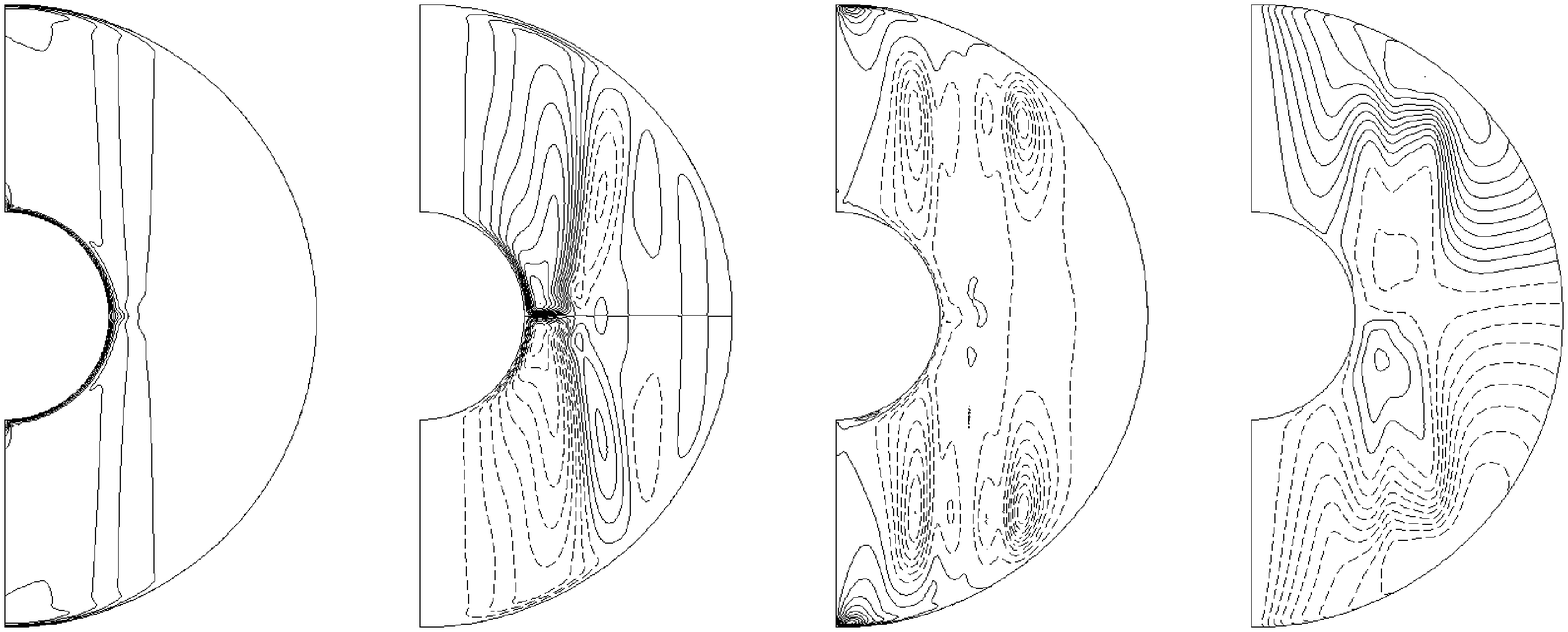}
\label{fig:fig2a}}
\subfigure[As in figure \ref{fig:fig1b}]
{\includegraphics[scale=0.4]{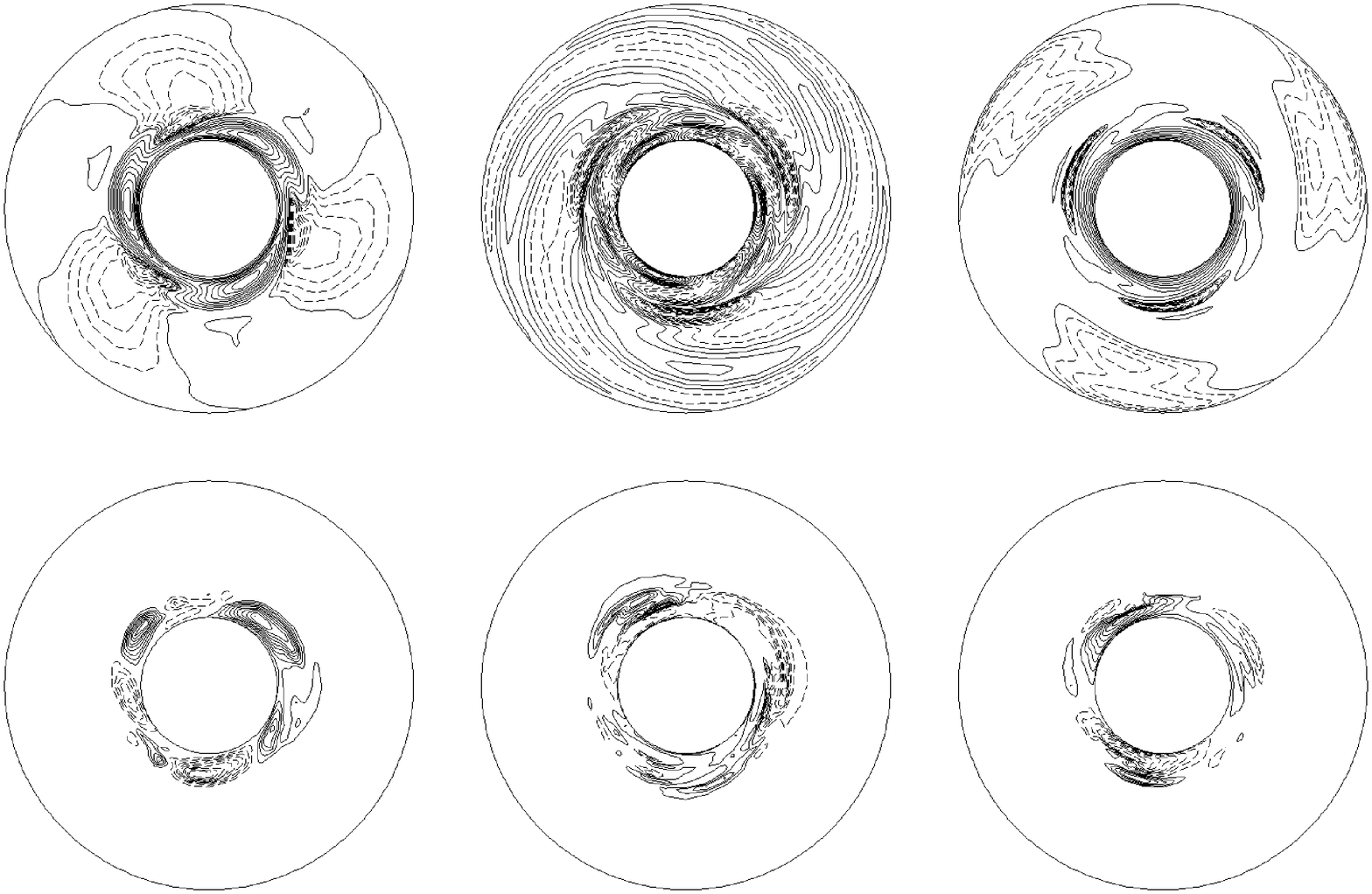}
\label{fig:fig2b}}
\caption{$E=10^{-3.5}$, $Ro=+1$ and $Rm_c=5000$.}
\label{fig:fig2}
\end{figure}

\begin{figure}[h]
\centering
\includegraphics[scale=0.9]{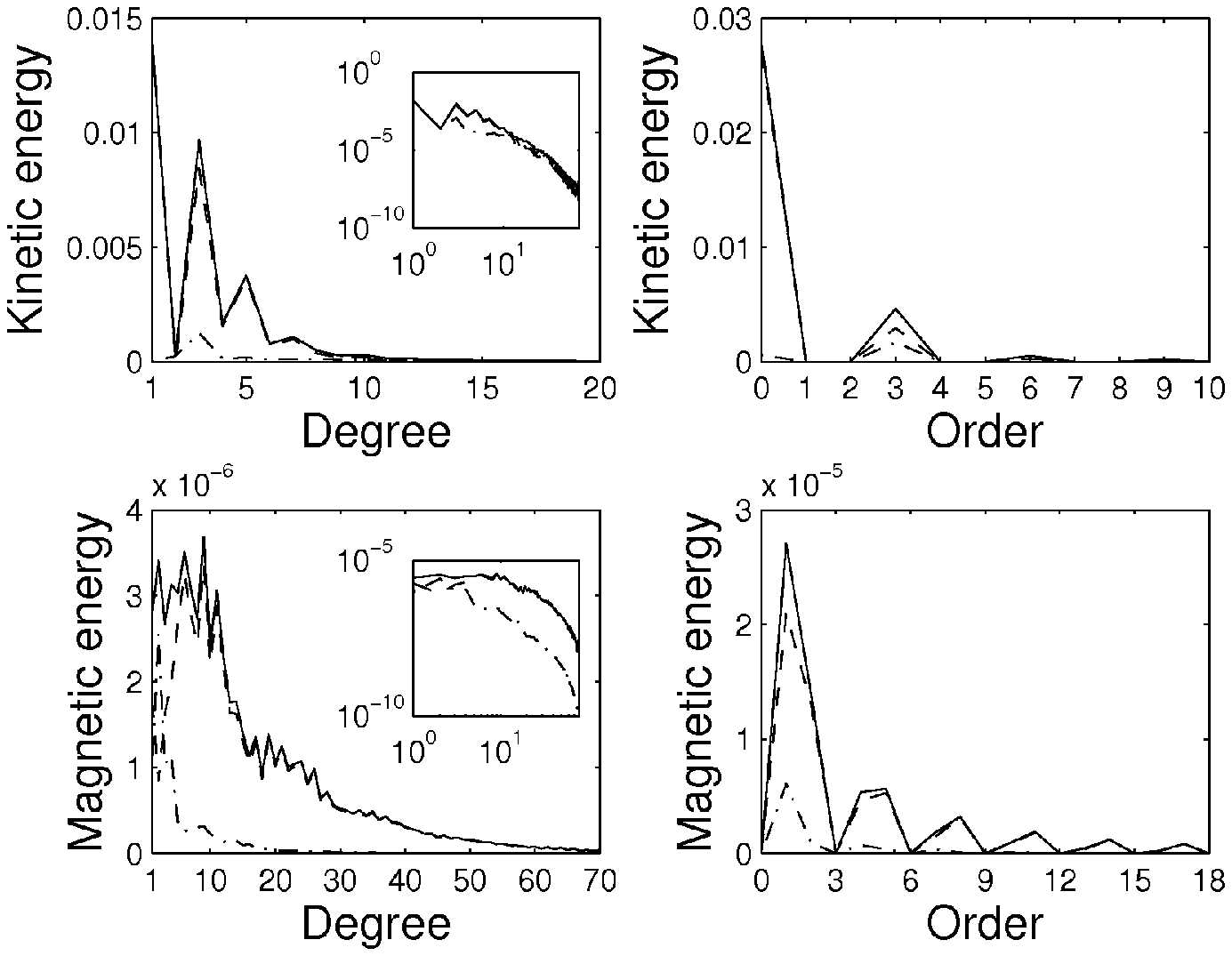}
\caption{As in figure \ref{fig:fig1c}, but for the solution in figure \ref{fig:fig2}. The $l=2$ contribution of flow is identically zero and thus is excluded from the log-log plot (inset figure in the upper left panel).}
\label{fig:fig2c}
\end{figure}

We then change the sign of $Ro$ to be positive, i.e. prograde flow. Figures \ref{fig:fig2} and \ref{fig:fig2c} show the dynamo solution at $E=10^{-3.5}$, $Ro=+1$ and $Rm_c=5000$. The critical $Rm$ for this prograde flow is greater than that for the retrograde flow. This asymmetry of the onset of dynamo action arises from the asymmetry between the prograde and retrograde flows, as we shall now illustrate.

Firstly, in the axisymmetric part of the two flows, we can observe that the Ekman layer of the $Ro=+1$ flow is thinner than that of the $Ro=-1$ flow (the first panels in figures \ref{fig:fig1a} and \ref{fig:fig2a}). When $|Ro|$ is very small, the thickness of the Ekman layer $\delta$ depends only on $E$, i.e. $\delta\sim\sqrt{E}$. This classic power law is derived from the balance among the viscous force, the Coriolis force and the pressure gradient with the neglect of the inertial force in the laminar Ekman layer (flows in our calculations are supercritical but still laminar). But when $|Ro|$ is finite the inertial force is not negligible and so $\delta$ also depends on $Ro$. The reason that $\delta$ of the $Ro=+1$ flow is thinner than that of the $Ro=-1$ flow is because of the average absolute rotation rate. The average absolute rotation rate of the $Ro=+1$ flow is between $+1$ and $+2$ whereas that of the $Ro=-1$ flow is between $+1$ and $0$. A stronger rotation rate induces a thinner Ekman layer. Moreover, the Ekman layer of the $Ro=+1$ flow is thickened near the equator whereas that of the $Ro=-1$ flow is squashed. This is because of the opposite directions of the meridional circulation (the second panels in figures \ref{fig:fig1a} and \ref{fig:fig2a}). Near the equator there is a radially outward jet in the $Ro=+1$ flow but a radially inward jet in the $Ro=-1$ flow.

Secondly, according to figure \ref{fig:fig0}, retrograde flow is more stable than prograde flow such that the $Ro=+1$ flow is more supercritical than the $Ro=-1$ flow ($Ro_c=+0.5$ for prograde flow and $Ro_c=-0.8$ for retrograde flow). This can be seen from the two flow patterns. In the meridional plane both the angular velocity and the meridional circulation of the $Ro=+1$ flow have more complex structures than those of the $Ro=-1$ flow. In the equatorial plane the $Ro=+1$ flow exhibits an $m=3$ mode which is consistent with the linear stability calculations for $E=10^{-3.5}$ (left panel in figure \ref{fig:fig0}). This mode occupies $14.7\%$ of the kinetic energy (upper right panel in figure \ref{fig:fig2c}) which is much larger than the $1.8\%$ of the $m=1$ mode in the $Ro=-1$ flow. Moreover, the spiral structure of the Rossby waves for prograde flow can be clearly perceived to spread radially outward (top row in figure \ref{fig:fig2b}) whereas the Rossby waves for retrograde flow are in the vicinity of the inner sphere (top row in figure \ref{fig:fig1b}).

All these differences between prograde and retrograde flows in respect of both axisymmetric and nonaxisymmetric parts can be mathematically interpreted through the Coriolis force. The Coriolis force is a first order term in the N-S equation such that it will bring opposite effects by changing the sign of the flow. Therefore it is not surprising that prograde flow requires a larger $Rm_c$ than retrograde flow, because in retrograde flow the $\alpha$ effect induced by the Rossby waves works near the tangent cylinder, where the $\omega$ effect induced by differential rotation works, but in prograde flow the Rossby waves spread radially outward and therefore the $\alpha$ and $\omega$ effects work at different locations, which weakens dynamo action.

The magnetic field in this solution has a rich spectrum governed by the selection rules of \citet{Bullard}. Since the flow is $m=3$ no harmonics of the form $m=3i\,(i=1,2,\ldots)$ can be generated, which is shown in the lower right panel in figure \ref{fig:fig2c}. This is more complex than the gently decaying spectrum for $Ro=-1$ shown in the lower right panel of figure \ref{fig:fig1c}, which is dominated by $m=0$ and $m=1$. The field has a quadrupolar symmetry about the equator. The equatorial symmetry depends on the parity of $(l-m)$ where $l$ and $m$ are the degree and order of the spherical harmonic components of the poloidal field. For the axisymmetric field, $m=0$ and then the equatorial symmetry is determined by $l$. In the $Ro=-1$ dynamo solution the dominant $l$ is odd, so the axisymmetric field is predominantly dipolar. However, in the $Ro=+1$ dynamo solution the dominant $l$ is even, so the axisymmetric field is predominantly quadrupolar.

Next we reduce the Ekman number to $10^{-4}$ and we firstly test retrograde flow. We try to use the flow at $Ro=-1$ to produce a dynamo, but unfortunately we fail to do so. Until $Rm=10^4$ we cannot find any dynamo. This is consistent with the nonlinear spherical Couette dynamo calculations \citep{Guervilly_Cardin} in which dynamo action in retrograde flow at $E=10^{-4}$ does not occur until $|Ro|$ is increased beyond $0.5$ which is translated to be $1.4$ in our definition. Here we should notice that $E$ in \citet{Guervilly_Cardin} and in our definition differs by a factor $(1-r_i/r_o)^2$ (equation (\ref{eq:gc}a)),
%{eq:ek_gc})), 
and so this comparison is not strictly correct, but it gives us a qualitative result that retrograde flow at low $E$ does not favour dynamo action. We may increase $|Ro|$ to be greater in a more supercritical regime, say, $Ro<-1$ in which the absolute rotation rate of inner sphere is in the opposite direction, but our computational power is not capable to do that.

\begin{figure}[h]
\centering
\subfigure[As in figure \ref{fig:fig1a}]
{\includegraphics[scale=0.4]{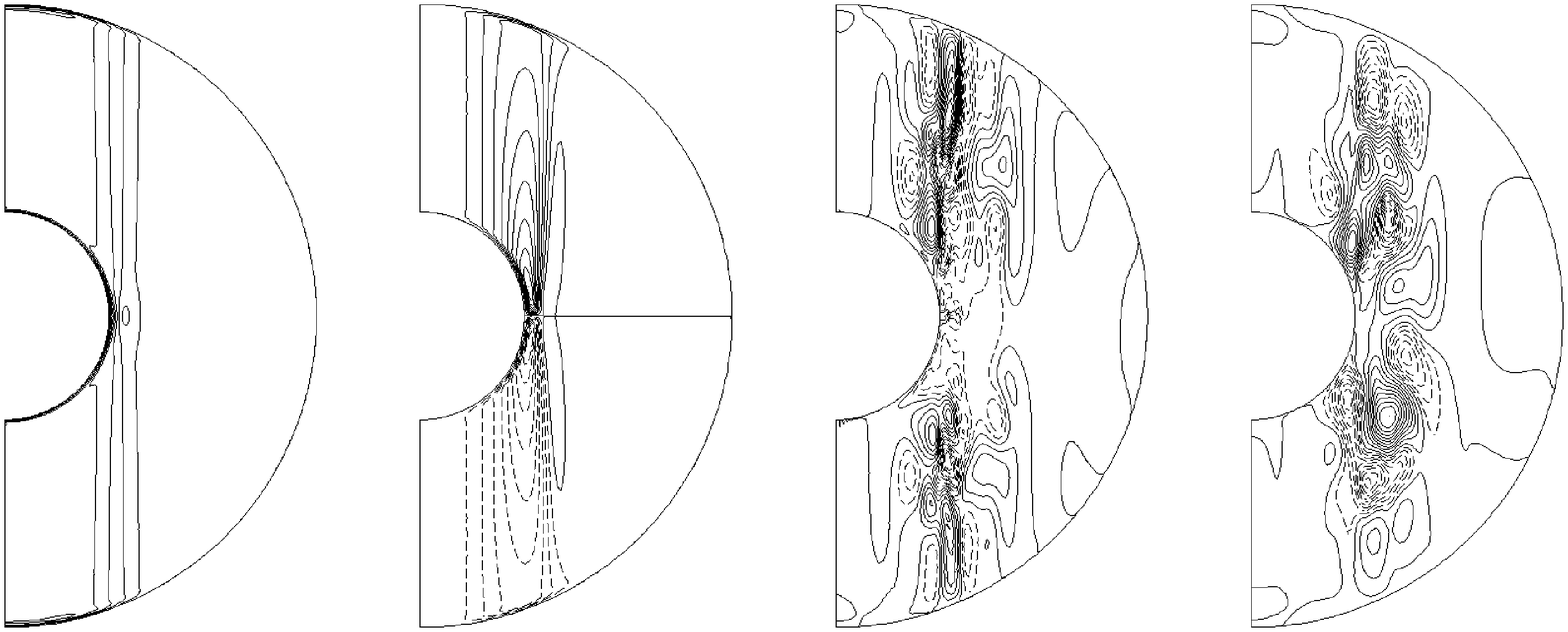}
\label{fig:fig3a}}
\subfigure[As in figure \ref{fig:fig1b}]
{\includegraphics[scale=0.4]{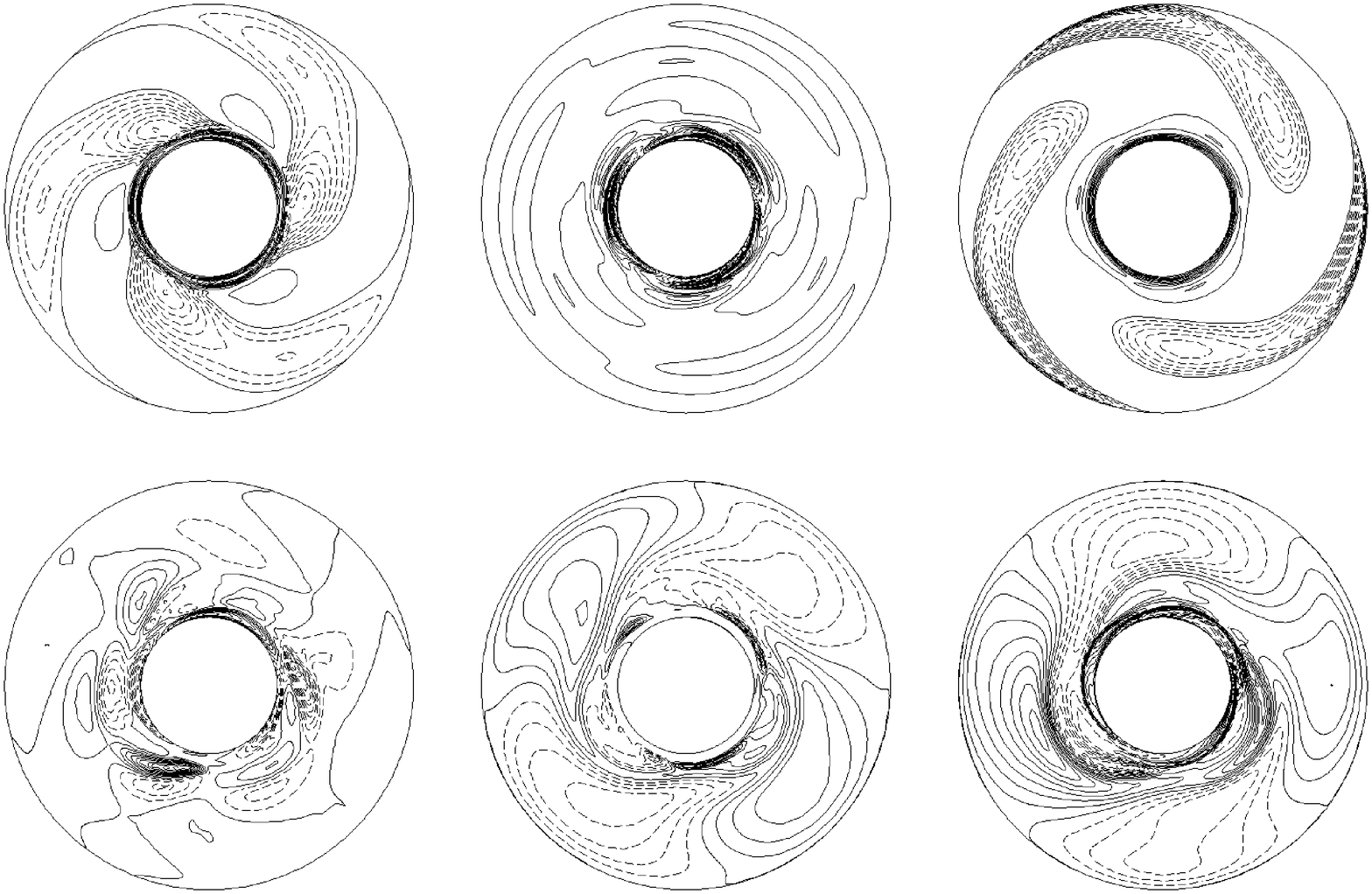}
\label{fig:fig3b}}
\caption{$E=10^{-4}$, $Ro=+0.3$ and $Rm_c=6500$.}
\label{fig:fig3}
\end{figure}

\begin{figure}[h]
\centering
\includegraphics[scale=0.9]{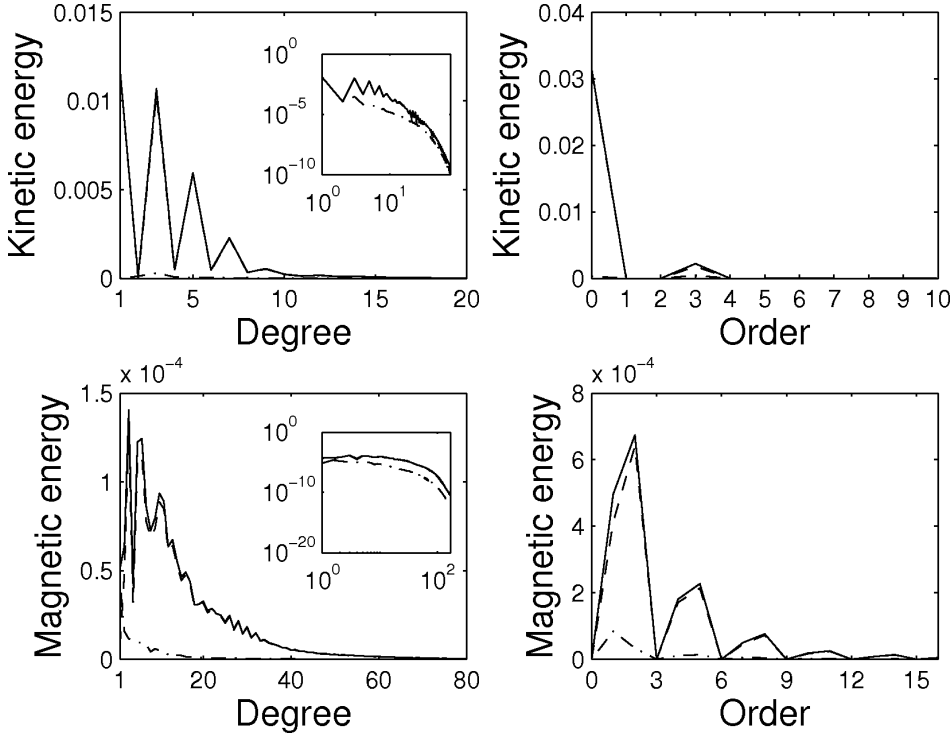}
\caption{As in figure \ref{fig:fig1c}, but for the solution in figure \ref{fig:fig3}. The $l=2$ contribution of flow is identically zero and thus is excluded from the log-log plot (inset figure in the upper left panel).}
\label{fig:fig3c}
\end{figure}

We now move to prograde flow. We choose $Ro$ to be $+0.3$ which is slightly above the neutral stability curve (left panel in figure \ref{fig:fig0}) and we find dynamos. Figures \ref{fig:fig3} and \ref{fig:fig3c} show the dynamo solution at $E=10^{-4}$, $Ro=+0.3$ and $Rm_c=6500$. Both the flow and field exhibit a more columnar structure than $E=10^{-3.5}$ (figures \ref{fig:fig3a} and \ref{fig:fig2a}) due to the stronger Coriolis effect (the Taylor-Proudman theorem). The flow exhibits a nice spiral structure and an $m=3$ Rossby wave spreads radially outward. The $m=2$ and $m=1$ modes of the magnetic field occupy, respectively, $40.0\%$ and $33.9\%$ of the total magnetic energy (lower right panel in figure \ref{fig:fig3c}) and the field has a quadrupolar symmetry because the dominant $l$ of the axisymmetric poloidal field is even.

The reason that at $E=10^{-4}$ a retrograde flow ($Ro=-1$) cannot easily produce a dynamo but a prograde flow ($Ro=+0.3$) can is because of the flow supercriticality. The asymmetry of linear stabilities between progarde and retrograde flows increases as the global rotation rate increases ($E$ decreases, see figure \ref{fig:fig0}). According to figure \ref{fig:fig0}, at $E=10^{-4}$ the magnitude of the critical $Ro$ of retrograde flow is $0.4$ ($Ro_c=-0.4$) but that of prograde flow is $0.2$ ($Ro_c=+0.2$). The former being twice as large as the latter, we may postulate that the flow supercriticality at $Ro=+0.3$ is already strong enough to generate a dynamo but the flow supercriticality at $Ro=-1$ is not yet sufficient. For example, the percentage of nonaxisymmetric modes in kinetic energy is $6.68\%$ for the $Ro=+0.3$ flow whereas it is $6.56\%$ for the $Ro=-1$ flow.

\begin{figure}[h]
\centering
\subfigure[As in figure \ref{fig:fig1a}]
{\includegraphics[scale=0.4]{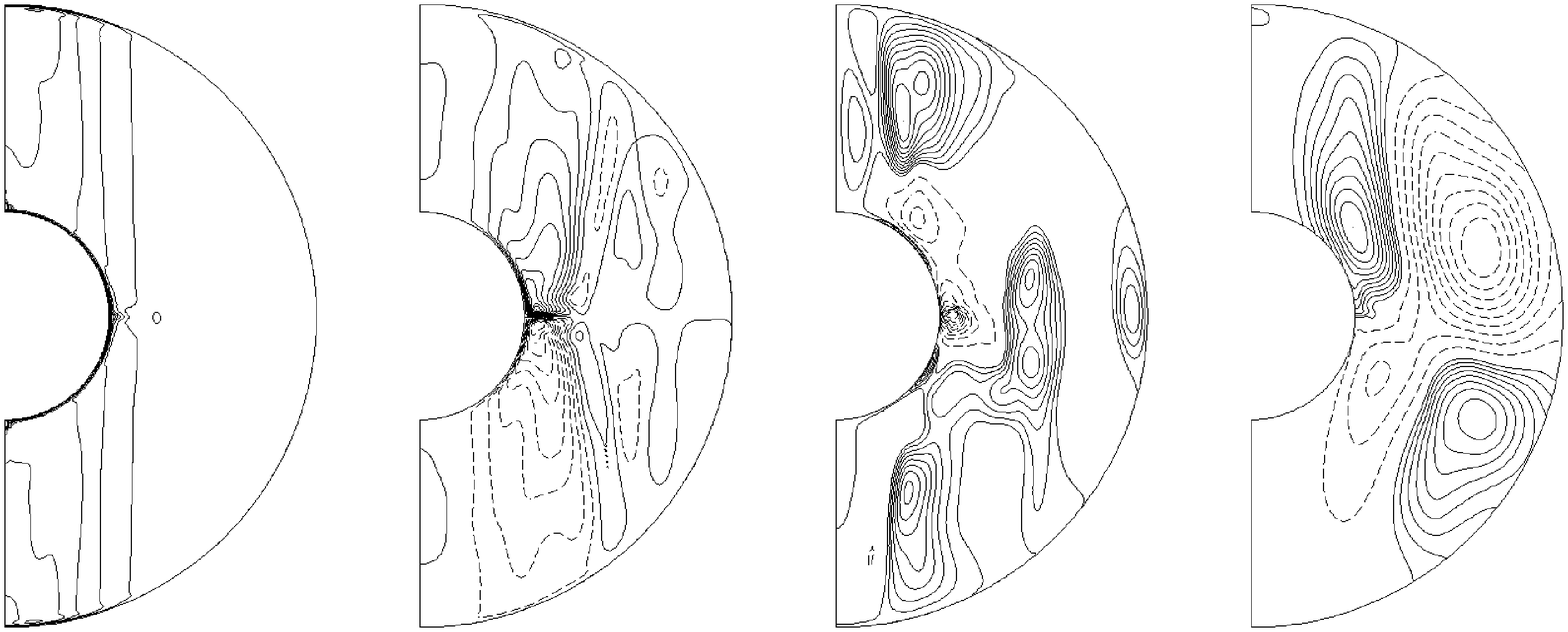}
\label{fig:fig4a}}
\subfigure[As in figure \ref{fig:fig1b}]
{\includegraphics[scale=0.4]{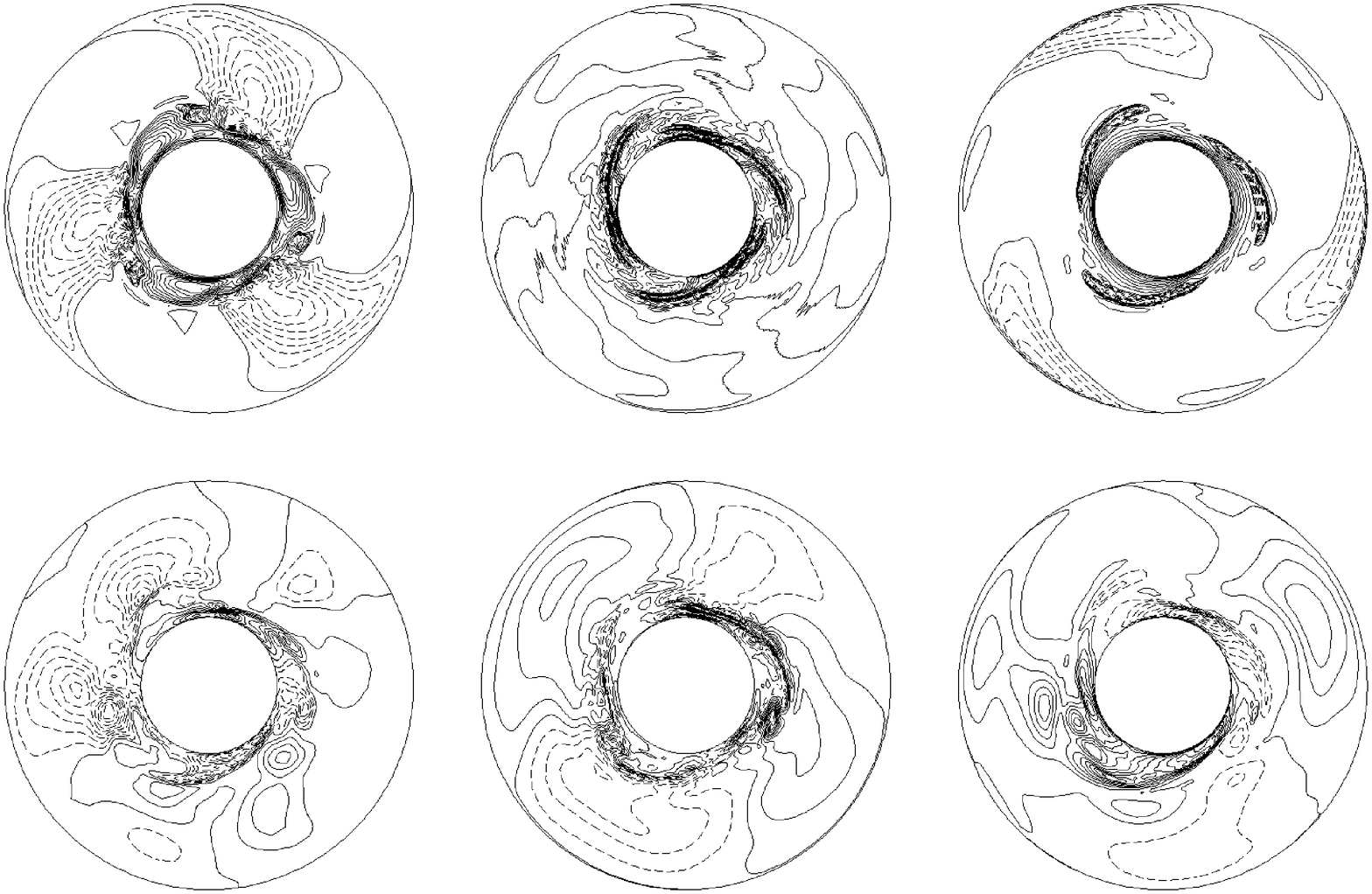}
\label{fig:fig4b}}
\caption{$E=10^{-4}$, $Ro=+1$ and $Rm_c=2000$.}
\label{fig:fig4}
\end{figure}

\begin{figure}[h]
\centering
\includegraphics[scale=0.9]{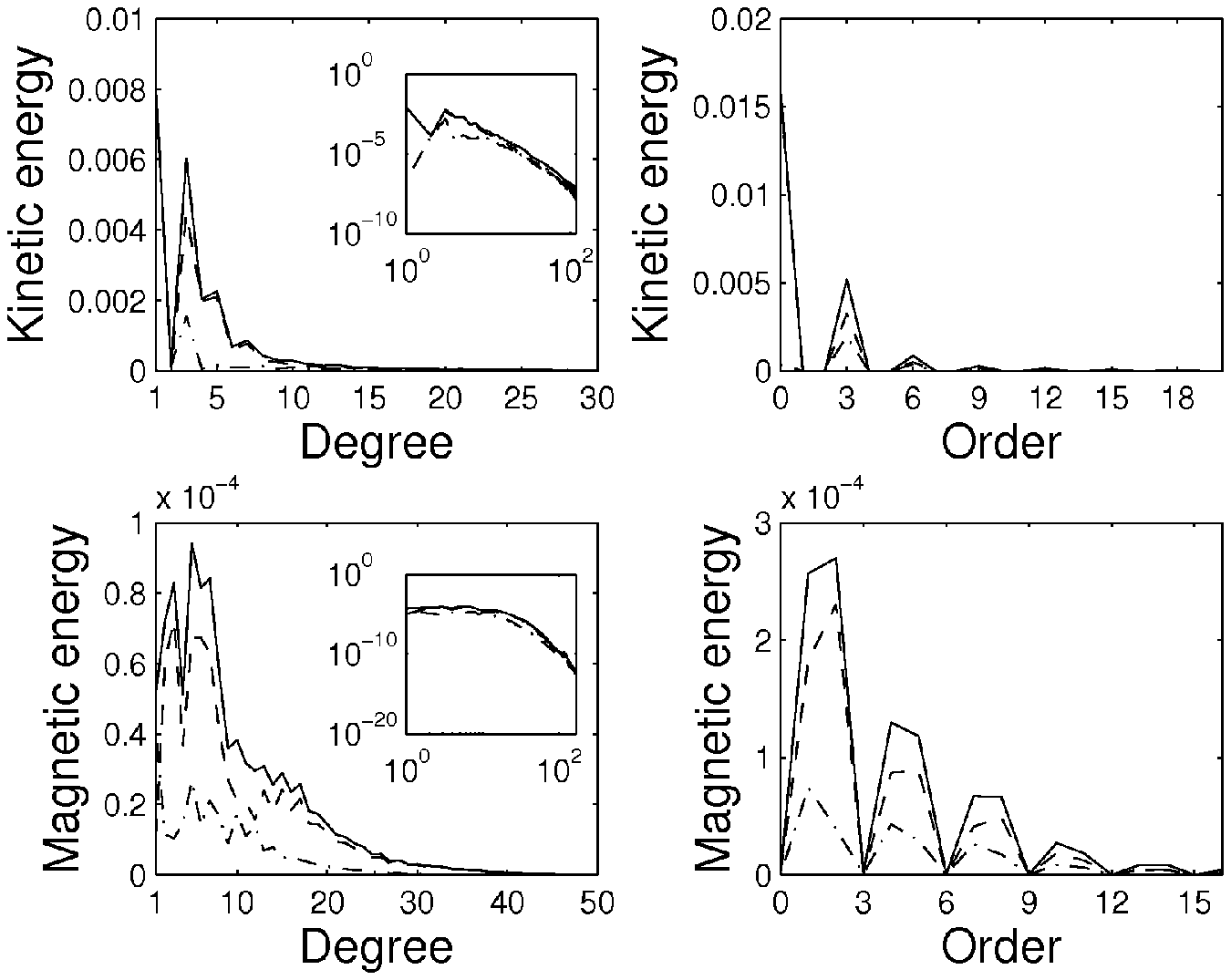}
\caption{As in figure \ref{fig:fig1c}, but for the solution in figure \ref{fig:fig4}. The $l=2$ contribution of flow is identically zero and thus is excluded from the log-log plot (inset figure in the upper left panel).}
\label{fig:fig4c}
\end{figure}

To verify this point that the strong flow supercriticality favours dynamo action, we continue to increase $Ro$ from $+0.3$ to $+1$. Figures \ref{fig:fig4} and \ref{fig:fig4c} show the dynamo solution at $E=10^{-4}$, $Ro=+1$ and $Rm_c=2000$. That $Rm_c$ at $Ro=+1$ is much lower than $Rm_c$ at $Ro=+0.3$ supports the point that the stronger flow supercriticality is better for dynamo action. On the other hand, comparison between $Rm_c=2000$ for $(E=10^{-4}, Ro=+1)$ and $Rm_c=5000$ for $(E=10^{-3.5}, Ro=+1)$ also supports this point. $Ro_c$ for $E=10^{-4}$ is $+0.2$ whereas $Ro_c$ for $E=10^{-3.5}$ is $+0.5$ (figure \ref{fig:fig0}), and then $Ro=+1$ results in stronger flow supercriticality for the $E=10^{-4}$ flow than for the $E=10^{-3.5}$ flow, such that $Rm_c$ of the $E=10^{-4}$ flow is lower than $Rm_c$ of the $E=10^{-3.5}$ flow.

Although the $Ro=+1$ flow also exhibits an $m=3$ mode (top row in figure \ref{fig:fig4b} and upper right panel in figure \ref{fig:fig4c}) as does the $Ro=+0.3$ flow, it has a more complex structure (top rows in figures \ref{fig:fig4b} and \ref{fig:fig3b}) arising from the stronger nonlinear inertial force. Accordingly the magnetic field has a more complex structure (bottom rows in figures \ref{fig:fig4b} and \ref{fig:fig3b}). Therefore we can conclude that the stronger flow supercriticality which leads to more complex structures in the flow and field facilitates the onset of dynamo action because of the more efficient interaction between different modes, namely the induction term ${\bm \nabla}\times\left({\bm U}\times {\bm B}\right)$ in the magnetic induction equation (\ref{eq:induction}). Although some simultations in turbulent regime showed that an increase of turbulence increased the dynamo threshold due to the enhancement of the magnetic diffusivity by the turbulence \citep{Bayliss}, our calculations in this supercritical regime are far away from the turbulent regime.

\begin{figure}[h]
\centering
\includegraphics[scale=0.4]{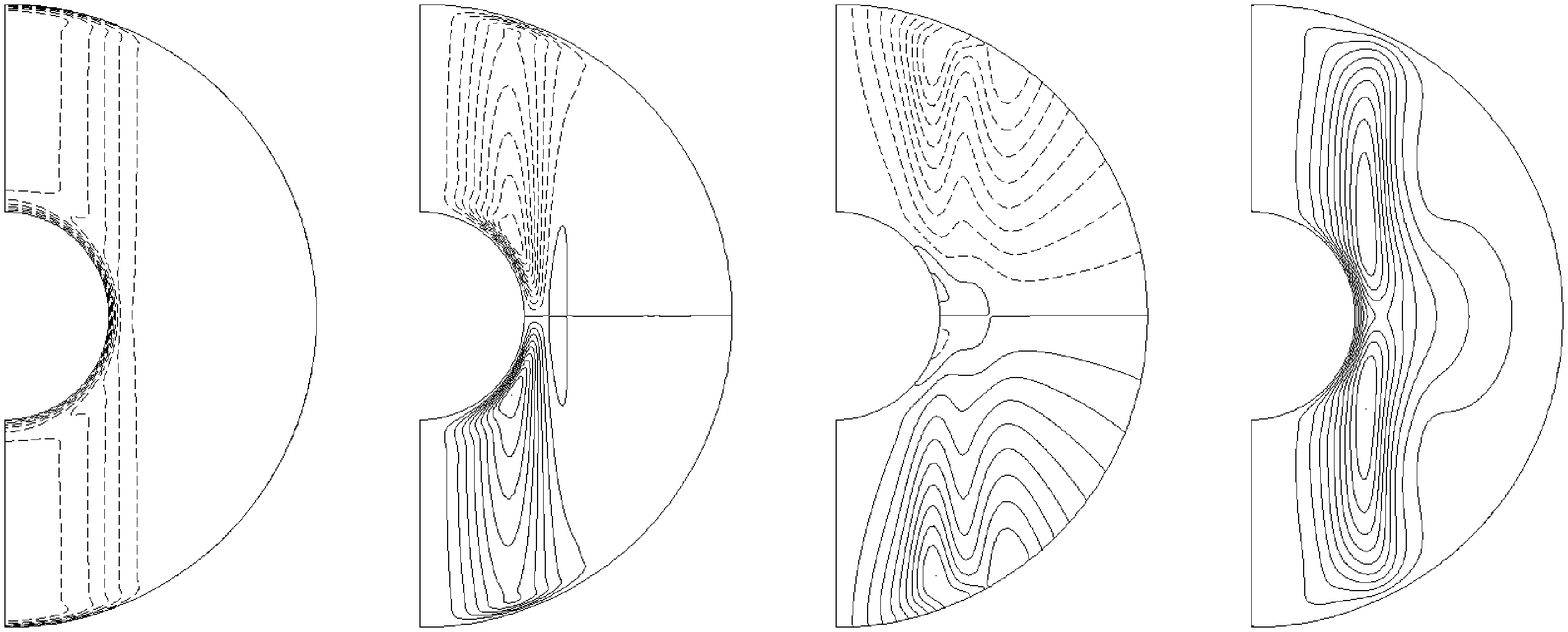}
\caption{$E=10^{-3.5}$, $Ro=-1$ and $Rm_c=2450$, and perfectly electrically conducting boundaries.}
\label{fig:fig5a}
\end{figure}

Following our calculations for different $E$ and $Ro$ ($Ro$ in respect of both sign and magnitude), we implement a perfectly electrically conducting boundary condition to investigate the influence of magnetic boundary conditions on dynamo action. Figure \ref{fig:fig5a} shows the dynamo solution at $E=10^{-3.5}$, $Ro=-1$ and $Rm_c=2450$ with both spheres being perfectly conducting. We show only the axisymmetric flow and field distributions in the meridional plane. The flow/field distribution in the equatorial plane is identical/similar to figure \ref{fig:fig1b} and so is not shown again. Compared with $Rm_c=2650$ for insulating boundaries in figure \ref{fig:fig1}, the conducting boundaries favour dynamo action, at least in this parameter regime of $E=10^{-3.5}$ and $Ro=-1$. This may be interpreted with the argument that in the case of conducting boundaries the poloidal field lines are trapped in the spherical shell and cannot penetrate out (the fourth panel in figure \ref{fig:fig5a}) such that no poloidal magnetic energy escapes, whereas the poloidal field lines with insulating boundaries can penetrate out (the fourth panel in figure \ref{fig:fig1a}) such that some poloidal magnetic energy is lost.

\begin{figure}[h]
\centering
\subfigure[As in figure \ref{fig:fig1a}]
{\includegraphics[scale=0.4]{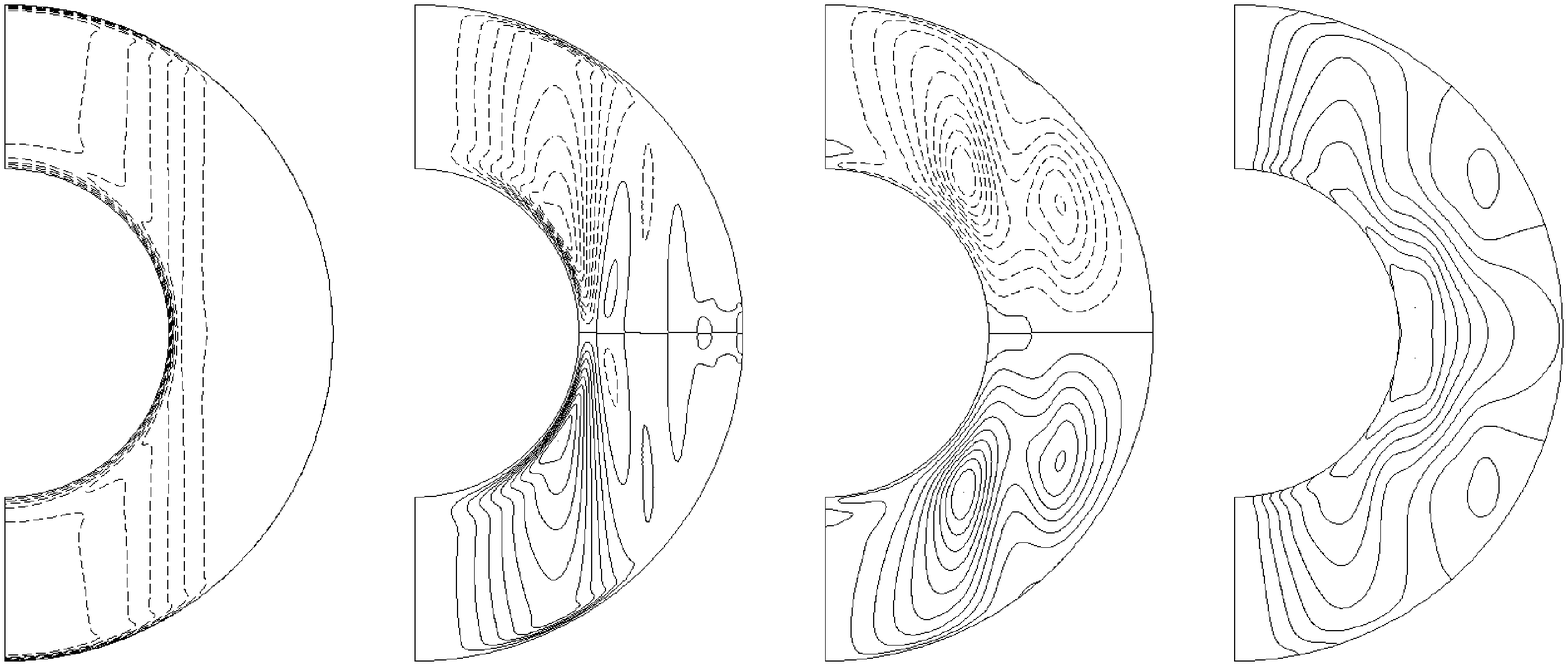}
\label{fig:fig6a}}
\subfigure[As in figure \ref{fig:fig1b}]
{\includegraphics[scale=0.4]{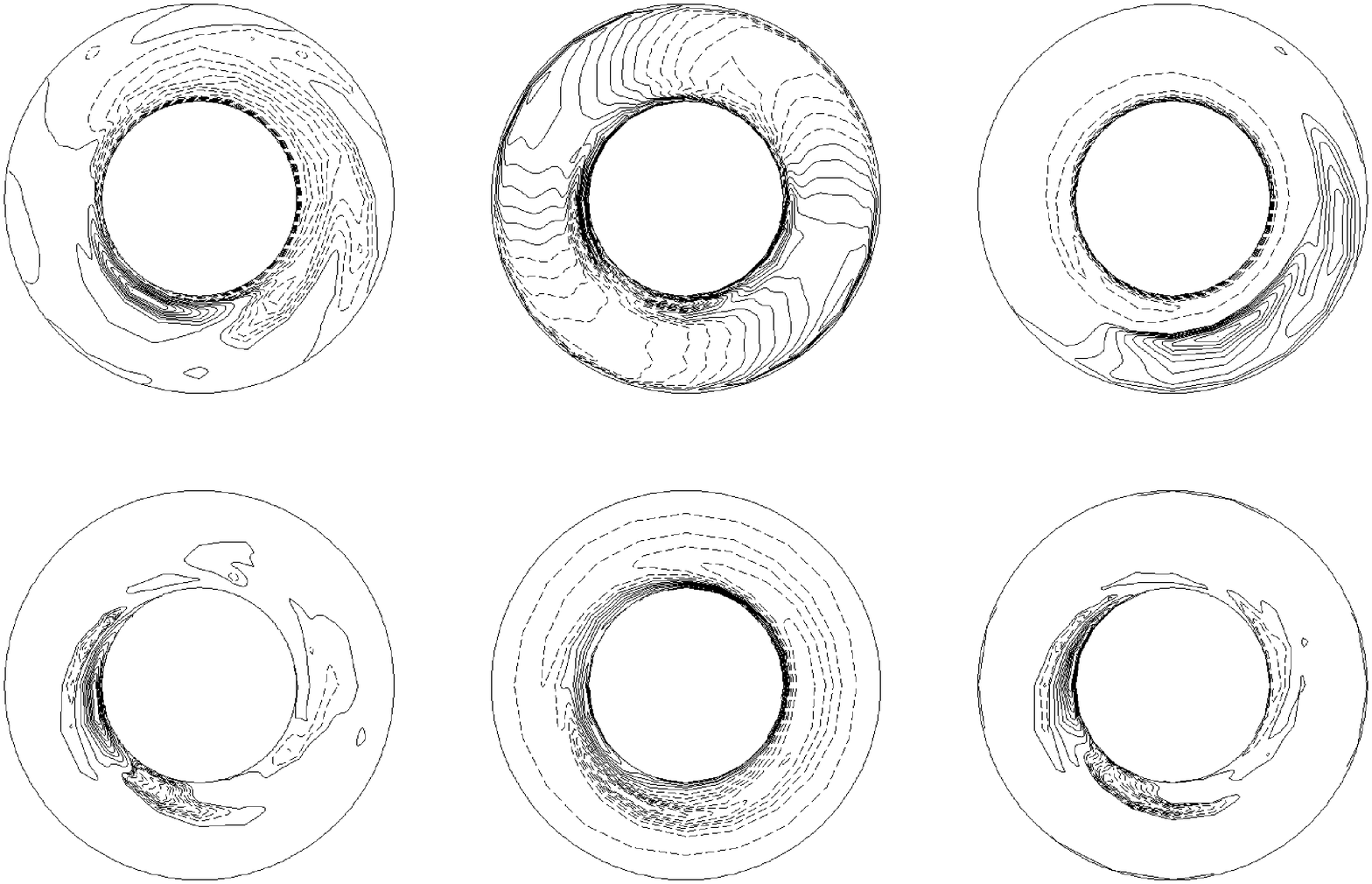}
\label{fig:fig6b}}
\caption{$E=10^{-3.5}$, $Ro=-1$ and $Rm_c=1250$, and aspect ratio $r_i/r_o=1/2$.}
\label{fig:fig6}
\end{figure}

In addition to the magnetic boundary condition, we perform calculations using a different aspect ratio $r_i/r_o=1/2$ to investigate the influence of the aspect ratio on dynamo action. Figure \ref{fig:fig6} shows the dynamo solution at $E=10^{-3.5}$, $Ro=-1$ and $Rm_c=1250$ with $r_i/r_o=1/2$. The flow exhibits an $m=1$ mode and the field has three comparable modes $m=0, 1$ and $2$. The critical magnetic Reynolds number $Rm_c=1250$ for $r_i/r_o=1/2$ is lower than $Rm_c=2650$ for $r_i/r_o=1/3$. This dynamo action is more efficient. One reason for this is that the total kinetic energy for $r_i/r_o=1/2$ is stronger than that for $r_i/r_o=1/3$. In our dimensionless equations the spherical gap $L$ is unity, and so the radii of the spheres are $r_o=1.5$ and $r_i=0.5$ for $r_i/r_o=1/3$, and $r_o=2$ and $r_i=1$ for $r_i/r_o=1/2$. According to equation (\ref{eq:u_bc}b), the inner boundary velocity for $r_i/r_o=1/2$ is twice that for $r_i/r_o=1/3$ such that a stronger flow is driven with $r_i/r_o=1/2$. Note that our current definition of the magnetic Reynolds number $Rm$ is based on the spherical shell width (equation (\ref{eq:parameters}c)). When one translates this into a new magnetic Reynolds number $Rm'$ based on the outer sphere radius given by
\begin{multiequations}
\singleequation
\begin{equation}\label{eq:rm'}
Rm'=\frac{|\varDelta\varOmega|\,r_o^2}{\eta},
\end{equation}
the following interrelation holds:
\begin{equation}\label{eq:rm_rm'}
Rm'=Rm\left(1-\frac{r_i}{r_o}\right)^{-2}.
\end{equation}
\end{multiequations}
Based on the outer shell magnetic Reynolds number, the critical $Rm'$ are $5960$ and $5000$ respectively for the cases $r_i/r_o=1/3$ and $r_i/r_o=1/2$. Thus one can argue that the small change in aspect ratio has led to a more efficient dynamo and a lower dynamo threshold. The reason for this increased efficiency is the strength of the $\alpha$ effect: in this dynamo solution with $r_i/r_o=1/2$, the percentages of kinetic energies in the $m=0$ and $m=1$ modes are, respectively, $96.1\%$ and $3.5\%$. Compared with the percentage of $1.8\%$ for the $m=1$ mode in the dynamo solution with $r_i/r_o=1/3$, the percentage of Rossby waves for the $r_i/r_o=1/2$ case is doubled such that the strength of the $\alpha$ effect is doubled. This point can be also supported by $Re_\omega$ and $Re_\alpha$ in Table \ref{tab:tab1}. In \citet{FDR} various aspect ratios for spherical Couette flow have been studied and the aspect ratio indeed alters the pattern of Rossby waves. This can be interpreted as a result of the $\beta$ effect, in which the curvature of the boundary induces Rossby waves. $\beta$ is defined as $\beta=1/h({\mathrm d}h/{\mathrm d}s)$ where $h$ is the height of the fluid column at cylindrical radius $s$. Clearly the value of $\beta$ is related to the aspect ratio. In the shallow water theory of a thin spherical gap, $\beta$ appears in the dispersion relationship for Rossby waves \citep{Pedlosky}, and although in the geometry of a wide spherical gap there is no explicit dispersion relationship, $\beta$ is still essential to Rossby waves. Therefore it is not surprising that $r_i/r_o=1/2$ is better for dynamo action because both the total flow and the $\alpha$ effect are stronger.

For a better understanding of the ``dynamo window'' in the nonlinear spherical Couette dynamo \citep{Guervilly_Cardin}, we test the corresponding kinematic spherical Couette dynamo. At first we translate the paramters of ``dynamo window'' to our definition and they are $E=10^{-2.6}$, $Ro=-4.3$ and $Rm_c\sim360$. Then we calculate with $E=10^{-2.6}$, $Ro=-4.3$ and $Rm=400$. However, we cannot find a dynamo. Therefore, we may speculate that this low $Rm$ ``dynamo window'' is mathematically a subcritical dynamo in the nonlinear regime.

To end this section we investigate the direction of magnetic dipole axis in all the six dynamo solutions. We compare between the contributions of two modes $(l=1,m=0)$ and $(l=1,m=1)$ of poloidal component, i.e. $h_{10}$ and $h_{11}$ (equations (\ref{eq:t_p}) and (\ref{eq:spherical})), to the radial field at the outer boundary. The former represents the axial dipole whereas the latter the equatorial dipole. It is interesting that the magnetic dipole is almost axial in the retrograde flows whereas it is almost equatorial in the prograde flows. It is not surprising because the magnetic field generated by the retrograde flows has very small portion of energy in the $m=1$ mode (lower right panel in figure \ref{fig:fig1c}) whereas the field generated by the prograde flows has very small portion of energy in the $m=0$ mode (lower right panels in figures \ref{fig:fig2c}, \ref{fig:fig3c} and \ref{fig:fig4c}). But we do not understand the physics why the direction of $\varDelta\varOmega$ is crucial to the direction of magnetic dipole axis.

\section{Conclusions}

In this work we have calculated examples of kinematic dynamos driven by spherical Couette flow. We draw the following conclusions.

2D axisymmetric flows at the Ekman number regime in our calculations cannot generate dynamo action whereas 3D flows can. This indicates that azimuthally drifting Rossby waves are crucial to dynamo action. The spherical Couette dynamos we have studied are $\alpha$-$\omega$ dynamos in which the $\omega$ effect is induced by differential rotation in the Stewartson layer and the $\alpha$ effect is induced by helical Rossby waves arising from the destabilisation of the Stewartson layer.

At a slow global rotation rate ($E=10^{-3.5}$), retrograde flow favours dynamo action but prograde flow does not, because in the former the $\alpha$ and $\omega$ effects work at the same location near the tangent cylinder but in the latter they two work at different locations. At a rapid global rotation rate ($E=10^{-4}$), prograde flow favours dynamo action but retrograde flow does not, because in the former the flow supercriticality arising from the nonlinear inertial force is strong enough to generate a dynamo whereas in the latter it is not sufficient. With a higher global rotation rate this asymmetry in the flow supercriticality is more prominent. Comparison between $Ro=+0.3$ and $+1$ at the same $E=10^{-4}$ and comparison between $E=10^{-3.5}$ and $10^{-4}$ at the same $Ro=+1$ both suggest that stronger flow supercriticality (leading to more complex structures of flow and field) facilitates the onset of dynamo action.

A conducting magnetic boundary condition and a larger aspect ratio both favour dynamo action. Through our calculation of kinematic spherical Couette dynamo we may speculate that the low critical $Rm$ ``dynamo window'' in the nonlinear spherical Couette dynamo could be mathematically interpreted as a subcritical dynamo. The magnetic dipole is almost axial in the retrograde flows and almost equatorial in the prograde flows. We also hope that this work may be of help to the ongoing experiments of spherical Couette dynamo, e.g. the dynamo experiment in Maryland \citep{Lathrop}.

\section*{Acknowledgements}
This work was partially supported by ERC grant 247303 (``MFECE'') to Andrew Jackson.

%\bibliographystyle{gGAF}
%\bibliography{paper}

\begin{thebibliography}{47}
\providecommand{\natexlab}[1]{#1}

\bibitem[\protect\citeauthoryear{Aurnou and Olson}{2001}]{Aurnou}
Aurnou, J.M. and Olson, P.L., Strong zonal winds from thermal convection in a
  rotating spherical shell. {\itshape Geophys. Res. Lett.} 2001, \textbf{28}, 2557--2559.

\bibitem[\protect\citeauthoryear{Avalos-Zuniga
  {\itshape{et~al.}}}{2009}]{Radler}
Avalos-Zuniga, R., Plunian, F. and R{\"a}dler, K.H., Rossby waves and $\alpha$
  effect. {\itshape Geophys. Astrophys. Fluid Dyn.} 2009, \textbf{103},
  375--396.

\bibitem[\protect\citeauthoryear{Bayliss {\itshape{et~al.}}}{2007}]{Bayliss}
Bayliss, R.A., Forest, C.B., Nornberg, M.D., Spence, E.J. and Terry, P.W.,
  Numerical simulations of current generation and dynamo excitation in a
  mechanically forced turbulent flow. {\itshape Phys. Rev. E} 2007,
  \textbf{75}, 026303.

\bibitem[\protect\citeauthoryear{Bullard and Gellman}{1954}]{Bullard}
Bullard, E.C. and Gellman, H., Homogeneous dynamos and terrestrial magnetism.
  {\itshape Proc. R. Soc. A} 1954, \textbf{247}, 213--278.

\bibitem[\protect\citeauthoryear{Busse}{1978}]{Busse}
Busse, Non-linear properties of thermal convection. {\itshape Rep. Prog. Phys.}
  1978, \textbf{41}, 1929–-1967.

\bibitem[\protect\citeauthoryear{Cardin and Brito}{2007}]{Grenoble}
Cardin, P. and Brito, D., Surveys of experimental dynamos. In {\itshape
  Mathematical Aspects of Natural Dynamos}, edited by E.~Dormy and A.~Soward,
  pp. 361--407 2007 (CRS Press: London).

\bibitem[\protect\citeauthoryear{Dormy {\itshape{et~al.}}}{1998}]{Dormy1}
Dormy, E., Cardin, P. and Jault, D., {MHD} flow in a slightly differentially
  rotating spherical shell with conducting inner core in a dipolar magnetic
  field. {\itshape Earth Planet. Sci. Lett.} 1998, \textbf{160}, 15--30.

\bibitem[\protect\citeauthoryear{Dormy {\itshape{et~al.}}}{2002}]{Dormy2}
Dormy, E., Jault, D. and Soward, A.M., A super-rotating shear layer in
  magnetohydrodynamic spherical {C}ouette flow. {\itshape J. Fluid Mech.} 2002,
  \textbf{452}, 263--291.

\bibitem[\protect\citeauthoryear{Galloway and Proctor}{1992}]{Galloway_Proctor}
Galloway, D.J. and Proctor, M.R.E., Numerical calculations of fast dynamos in
  smooth velocity fields with realistic diffusion. {\itshape Nature} 1992,
  \textbf{356}, 691--693.

\bibitem[\protect\citeauthoryear{Gubbins}{2008}]{Gubbins}
Gubbins, D., Implication of kinematic dynamo studies for the geodynamo.
  {\itshape Geophys. J. Int.} 2008, \textbf{173}, 79--91.

\bibitem[\protect\citeauthoryear{Gubbins
  {\itshape{et~al.}}}{2000{\natexlab{a}}}]{Gubbins1}
Gubbins, D., Barber, C.N., Gibbons, S. and Love, J.J., Kinematic dynamo action
  in a sphere. {I}. {E}ffects of differential rotation and meridional
  circulation on solutions with axial dipole symmetry. {\itshape Proc. R. Soc.
  A} 2000{\natexlab{a}}, \textbf{456}, 1333--1353.

\bibitem[\protect\citeauthoryear{Gubbins
  {\itshape{et~al.}}}{2000{\natexlab{b}}}]{Gubbins2}
Gubbins, D., Barber, C.N., Gibbons, S. and Love, J.J., Kinematic dynamo action
  in a sphere. {II}. {S}ymmetry selection. {\itshape Proc. R. Soc. A}
  2000{\natexlab{b}}, \textbf{456}, 1669--1683.

\bibitem[\protect\citeauthoryear{Gubbins and Gibbons}{2009}]{Gubbins_Gibbons}
Gubbins, D. and Gibbons, S.J., Kinematic dynamo action in a sphere with weak
  differential rotation. {\itshape Geophys. J. Int.} 2009, \textbf{177},
  71--81.

\bibitem[\protect\citeauthoryear{Guervilly and Cardin}{2010}]{Guervilly_Cardin}
Guervilly, C. and Cardin, P., Numerical simulations of dynamos generated in
  spherical {C}ouette flows. {\itshape Geophys. Astrophys. Fluid Dyn.} 2010,
  \textbf{104}, 221--248.

\bibitem[\protect\citeauthoryear{Hollerbach}{2000{\natexlab{a}}}]{LNP}
Hollerbach, R., Magnetohydrodynamic flows in spherical shells. In {\itshape
  Physics of Rotating Fluids (Lecture Notes in Physics)}, edited by C.~Egbers
  and G.~Pfister, pp. 295--316 2000 (Springer: Heidelberg).

\bibitem[\protect\citeauthoryear{Hollerbach}{2000{\natexlab{b}}}]{IJNMF}
Hollerbach, R., A spectral solution of the magneto-convection equations in
  spherical geometry. {\itshape Int. J. Numer. Meth. Fluids}
  2000{\natexlab{b}}, \textbf{32}, 773--797.

\bibitem[\protect\citeauthoryear{Hollerbach}{2003}]{JFM}
Hollerbach, R., Instabilities of the {S}tewartson layer. {P}art 1. {T}he
  dependence on the sign of {$Ro$}. {\itshape J. Fluid Mech.} 2003,
  \textbf{492}, 289--302.

\bibitem[\protect\citeauthoryear{Hollerbach}{2009}]{PRSA2}
Hollerbach, R., Non-axisymmetric instabilities in magnetic spherical {C}ouette
  flow. {\itshape Proc. R. Soc. A} 2009, \textbf{465}, 2003--2013.

\bibitem[\protect\citeauthoryear{Hollerbach {\itshape{et~al.}}}{2007}]{EJMB}
Hollerbach, R., Canet, E. and Fournier, A., Spherical {C}ouette flow in a
  dipolar magnetic field. {\itshape European. J. Mech. B} 2007, \textbf{26},
  729--737.

\bibitem[\protect\citeauthoryear{Hollerbach {\itshape{et~al.}}}{2004}]{TCFD}
Hollerbach, R., Futterer, B., More, T. and Egbers, C., Instabilities of the
  {S}tewartson layer. {P}art 2. {S}upercritical mode transitions. {\itshape
  Theoret. Comput. Fluid Dynamics} 2004, \textbf{18}, 197--204.

\bibitem[\protect\citeauthoryear{Hollerbach {\itshape{et~al.}}}{2006}]{FDR}
Hollerbach, R., Junk, M. and Egbers, C., Non-axisymmetric instabilities in
  basic state spherical {C}ouette flow. {\itshape Fluid Dyn. Res.} 2006,
  \textbf{38}, 257--273.

\bibitem[\protect\citeauthoryear{Hollerbach and Skinner}{2001}]{PRSA1}
Hollerbach, R. and Skinner, S., Instabilities of magnetically induced shear
  layers and jets. {\itshape Proc. R. Soc. A} 2001, \textbf{457}, 785--802.

\bibitem[\protect\citeauthoryear{Jones}{2011}]{Jones}
Jones, C.A., Planetary magnetic fields and fluid dynamos. {\itshape Annu. Rev.
  Fluid Mech.} 2011, \textbf{43}, 583--614.

\bibitem[\protect\citeauthoryear{Krause and R{\"a}dler}{1980}]{meanfield}
Krause, F. and R{\"a}dler, K.H., chapter 1. In {\itshape Mean-field
  Magnetohydrodynamics and Dynamo Theory} 1980 (Pergamon Press: Oxford).

\bibitem[\protect\citeauthoryear{Livermore
  {\itshape{et~al.}}}{2007}]{Livermore}
Livermore, P.W., Hughes, D.W. and Tobias, S.M., The role of helicity and
  stretching in forced kinematic dynamos in a spherical shell. {\itshape Phys.
  Fluids} 2007, \textbf{19}, 057101.

\bibitem[\protect\citeauthoryear{Manneville and Olson}{1996}]{Manneville}
Manneville, J.B. and Olson, P., Banded convection in rotating fluid spheres and
  the circulation of the jovian atmosphere. {\itshape Icarus} 1996,
  \textbf{122}, 242--250.

\bibitem[\protect\citeauthoryear{Moffatt}{1978}]{Moffatt}
Moffatt, H.K., chapter 10. In {\itshape Magnetic Field Generation in
  Electrically Conducting Fluids} 1978 (Cambridge University Press: Cambridge).

\bibitem[\protect\citeauthoryear{Nataf {\itshape{et~al.}}}{2008}]{Grenoble2}
Nataf, H.C., Alboussi{\'e}re, T., Brito, D., Cardin, P., Gagni{\'e}re, N.,
  Jault, D. and Schmitt, D., Rapidly rotating spherical {C}ouette flow in a
  dipolar magnetic field: an experimental study of the mean axisymmetric flow.
  {\itshape Phys. Earth Planet. Inter.} 2008, \textbf{170}, 60--72.

\bibitem[\protect\citeauthoryear{Parker}{1955}]{Parker}
Parker, E.N., Hydromagnetic dynamo models. {\itshape Astrophys. J.} 1955,
  \textbf{122}, 293--314.

\bibitem[\protect\citeauthoryear{Pedlosky}{1987}]{Pedlosky}
Pedlosky, J., chapter 3. In {\itshape Geophysical Fluid Dynamics (second
  edition)} 1987 (Springer: Heidelberg).

\bibitem[\protect\citeauthoryear{Peffley {\itshape{et~al.}}}{2000}]{Lathrop}
Peffley, N.L., Goumilevski, A.G., Cawthrone, A.B. and Lathrop, D.P.,
  Characterization of experimental dynamos. {\itshape Geophys. J. Int.} 2000,
  \textbf{142}, 52--58.

\bibitem[\protect\citeauthoryear{Roberts}{1972}]{Roberts}
Roberts, P.H., Kinematic dynamo models. {\itshape Phil. Trans. R. Soc. A} 1972,
  \textbf{272}, 663--698.

\bibitem[\protect\citeauthoryear{Roberts and
  Glatzmaier}{2000}]{Roberts_Glatzmaier}
Roberts, P.H. and Glatzmaier, G.A., Geodynamo theory and simulations. {\itshape
  Rev. Mod. Phys.} 2000, \textbf{72}, 1081--1123.

\bibitem[\protect\citeauthoryear{R{\"u}diger}{1989}]{Ruediger}
R{\"u}diger, G., chapter 4. In {\itshape Differential rotation and stellar
  convection} 1989 (Akademia-Verlag: Berlin).

\bibitem[\protect\citeauthoryear{R{\"u}diger and
  Hollerbach}{2005}]{Ruediger_Hollerbach}
R{\"u}diger, G. and Hollerbach, R., chapter 1. In {\itshape The Magnetic
  Universe: Geophysical and Astrophysical Dynamo Theory} 2005 (Wiley-VCH press:
  Berlin).

\bibitem[\protect\citeauthoryear{Sakuraba and Roberts}{2009}]{Sakuraba}
Sakuraba, A. and Roberts, P.H., Generation of a strong magnetic field using
  uniform heat flux at the surface of the core. {\itshape Nature Geoscience}
  2009, \textbf{2}, 802--805.

\bibitem[\protect\citeauthoryear{Schaeffer and
  Cardin}{2005}]{Schaeffer_Cardin1}
Schaeffer, N. and Cardin, P., Quasi-geostrophic model of the instabilities of
  the {S}tewartson layer in flat and depth-varying containers. {\itshape Phys.
  Fluids} 2005, \textbf{17}, 104111.

\bibitem[\protect\citeauthoryear{Schaeffer and Cardin}{2006}]{Schaeffer_Cardin}
Schaeffer, N. and Cardin, P., Quasi-geostrophic kinematic dynamos at low
  magnetic prandtl number. {\itshape Earth Planet. Sci. Lett.} 2006,
  \textbf{245}, 595--604.

\bibitem[\protect\citeauthoryear{Schmitt {\itshape{et~al.}}}{2008}]{Grenoble1}
Schmitt, D., Alboussi{\'e}re, T., Brito, D., Cardin, P., Gagni{\'e}re, N.,
  Jault, D. and Nataf, H.C., Rotating spherical {C}ouette flow in a dipolar
  magnetic field: Experimental evidence of hydromagnetic waves. {\itshape J.
  Fluid Mech.} 2008, \textbf{604}, 175--197.

\bibitem[\protect\citeauthoryear{Sisan {\itshape{et~al.}}}{2004}]{Lathrop1}
Sisan, D.R., Mujica, N., Tillotson, W.A., Huang, Y.M., Dorland, W., Hassam,
  A.B., Antonsen, T.M. and Lathrop, D.P., Experimental observation and
  characterization of the magnetorotational instability. {\itshape Phys. Rev.
  Lett.} 2004, \textbf{93}, 114502.

\bibitem[\protect\citeauthoryear{Souriau and Poupinet}{2003}]{Souriau}
Souriau, A. and Poupinet, G., Inner core rotation: a critical appraisal. In
  {\itshape Earth's {C}ore: {D}ynamics, {S}tructure, {R}otation (Geodynamics
  Series 31)}, edited by V.~Dehant, K.C. Creager, S.~Karato and S.~Zatman, pp.
  65--82 2003 (American Geophysical Union: Washington).

\bibitem[\protect\citeauthoryear{Starchenko}{1998}]{Starchenko}
Starchenko, S.V., Magnetohydrodynamic flow between insulating shells rotating
  in strong potential field. {\itshape Phys. Fluids} 1998, \textbf{10},
  2412--2420.

\bibitem[\protect\citeauthoryear{Stewartson}{1966}]{Stewartson}
Stewartson, K., On almost rigid rotations. {P}art 2. {\itshape J. Fluid Mech.}
  1966, \textbf{26}, 131--144.

\bibitem[\protect\citeauthoryear{Wei and Hollerbach}{2008}]{PRE}
Wei, X. and Hollerbach, R., Instabilities of {S}hercliff and {S}tewartson
  layers in spherical {C}ouette flow. {\itshape Phys. Rev. E} 2008,
  \textbf{78}, 026309.

\bibitem[\protect\citeauthoryear{Wicht and Tilgner}{2010}]{Wicht_Tilgner}
Wicht, J. and Tilgner, A., Theory and modeling of planetary dynamos. {\itshape
  Space Sci. Rev.} 2010, \textbf{152}, 501--542.

\bibitem[\protect\citeauthoryear{Willis and Barenghi}{2002}]{Willis_Barenghi}
Willis, A.P. and Barenghi, C.F., A {T}aylor-{C}ouette dynamo. {\itshape Astron.~Astrophys}
  2002, \textbf{393}, 339--343.

\bibitem[\protect\citeauthoryear{Zhang {\itshape{et~al.}}}{2005}]{Zhang_Song}
Zhang, J., Song, X., Li, Y., Richards, P.G., Sun, X. and Waldhauser, F., Inner
  core differential motion confirmed by earthquake doublet waveform doublets.
  {\itshape Science} 2005, \textbf{309}, 1357--1360.

\end{thebibliography}

\end{document}